\documentclass[%
 reprint,
 amsmath,amssymb,
 aps,
]{revtex4-1}

\usepackage{graphicx}
\usepackage{psfrag}
\usepackage{dcolumn}
\usepackage{bm}

\usepackage{hyperref} 
\usepackage{xcolor}
\hypersetup{
    colorlinks,
    linkcolor={red!50!black},
    citecolor={blue!70!black},
    urlcolor={blue!90!black}
}
\begin{document}

\preprint{APS/123-QED}

\title{Nanoscale dynamics during self-organized ion beam patterning of Si:\\II. Kr$^+$ Bombardment}

\author{Peco Myint}
 \email{peco@bu.edu}
\affiliation{%
Division of Materials Science and Engineering, \\Boston University, Boston, Massachusetts 02215 USA
\\ \& X-ray Science Division, Argonne National Laboratory, 
\\Lemont, Illinois 60439, USA}

\author{Karl F. Ludwig, Jr.}%
 \email{ludwig@bu.edu}
\affiliation{%
Department of Physics and\\Division of Materials Science and Engineering, \\Boston University, Boston, Massachusetts 02215 USA}

\author{Lutz Wiegart}
\author{Yugang Zhang}
\author{Andrei Fluerasu}
\affiliation{
 National Synchrotron Light Source II, \\Brookhaven National Lab, Upton, NY 11973 USA
}%

\author{Xiaozhi Zhang}
\author{Randall L. Headrick}
\affiliation{
 Department of Physics and Materials Science Program, \\University of Vermont, Burlington, Vermont 05405 USA
}%

\date{\today}

\begin{abstract}
{\noindent Understanding the self-organized ion beam nanopatterning of elemental semiconductors, particularly silicon, is of intrinsic scientific and technological interest.  This is the second component of a two-part coherent x-ray scattering and X-ray Photon Correlation Spectroscopy (XPCS) investigation of the kinetics and fluctuation dynamics of nanoscale ripple development on silicon during 1 keV Ar$^+$ (Part I \cite{myint12020nanoscale}) and Kr$^+$ bombardment at 65 $^{\circ}$ polar angle.  Here, it's found that the ion-enhanced viscous flow relaxation is essentially equal for Kr$^+$-induced patterning as previously found for Ar$^+$ patterning despite the difference in ion masses.  However, the magnitude of the surface curvature-dependent roughening rate in the early-stage kinetics is larger for Kr$^+$ than for Ar$^+$, consistent with expectations that the heavier ion gives an increased mass redistributive contribution to the initial surface instability.  As with the Ar$^+$ case, fluctuation dynamics in the late stage show a peak in correlation times at the length scale corresponding to the dominant structural feature on the surface - the ripples.  Finally, it's shown that speckle motion during the surface evolution can be analyzed to determine spatial inhomogeneities in erosion rate and ripple velocity.  This allows the direction and speed of ripple motion to be measured in a real time experiment.  In the present case, ripple motion is found to be into the projected direction of the ion source, in contrast to expectations from an existing sputter erosion driven model with parameters derived from binary collision approximation simulations.}

\end{abstract}

\maketitle

\section{\label{sec:level1}Introduction}
Broad beam low energy ion bombardment of surfaces can lead to the self-organized formation of a remarkable range of patterns including nanodots \cite{ozaydin2008effects}, nanoscale ripples \cite{chan2007making} and nanoscale pits/holes \cite{wei2009self}, as well as to ultrasmoothening \cite{moseler2005ultrasmoothness}.  Understanding the nanopatterning of elemental semiconductors, particularly silicon, is both foundational for the broader field and of intrinsic scientific and technological interest itself. In the case of elemental semiconductors patterned at room temperature, the surface is amorphized by the ions and off-axis bombardment can produce ripple patterns.  Competing theories of self-organized ion beam nanopatterning advocate for or combine models of different physical processes believed to play important roles, including curvature-dependent sputtering  \cite{sigmund1969theory,sigmund1973mechanism,bradley1988theory}, lateral mass redistribution \cite{carter1996roughening}, surface diffusion \cite{bradley1988theory}, ion-enhanced viscous flow \cite{umbach2001spontaneous} and stress-induced flow \cite{castro2012hydrodynamic,castro2012stress,norris2012stress,moreno2015nonuniversality,munoz2019stress}.  Despite extensive study, there is still not agreement on which processes are dominating surface evolution in a given situation. 

This is the second component of a two-part investigation using X-ray Photon Correlation Spectroscopy (XPCS) to examine the evolution of average surface kinetics and surface fluctuation dynamics during room temperature nanopatterning of silicon by the two most widely used inert gas ions: Ar$^+$ (Part I) and Kr$^+$ (the present work).  In this Part, we compare observed kinetics and dynamics to that reported for Ar$^+$ ion nanopatterning in Part I in order to investigate the effect of mass within a consistent experimental framework capable of measuring the coefficients entering linear theory formalisms.  As noted in Part I, the application of XPCS studies to surface processes is still relatively rare and an important goal of this work is also to explore the capabilities of the technique.  In this Part we implement an XPCS approach which allows us to measure in real time the direction and speed of propagation of the self-organized ripples on the surface, driven by continued bombardment.  The ripple velocity is a fundamental parameter predicted by competing theories, but its measurement has previously been challenging.

The plan of the paper is as follows: In Sect. II below we describe the methods used in the experiments and simulations.  Section III provides a broad brush overview of the basic behavior observed in the speckle-averaged x-ray intensity evolution during nano-ripple formation.  The early stages of the patterning process are analyzed in Sect. IV to determine the linear theory coefficients which are compared with theoretical predictions and with results from Ar$^+$ nanopatterning.  Section V examines the late-stage coarsening kinetics in both experiment and simulation while the evolution of fluctuation dynamics is examined in Sect. VI.  Section VII presents the measurement of ripple velocity while the results and their implications are discussed in Sect. VIII.

\section{Methods}

\begin{figure}
\includegraphics[width=3.2 in]{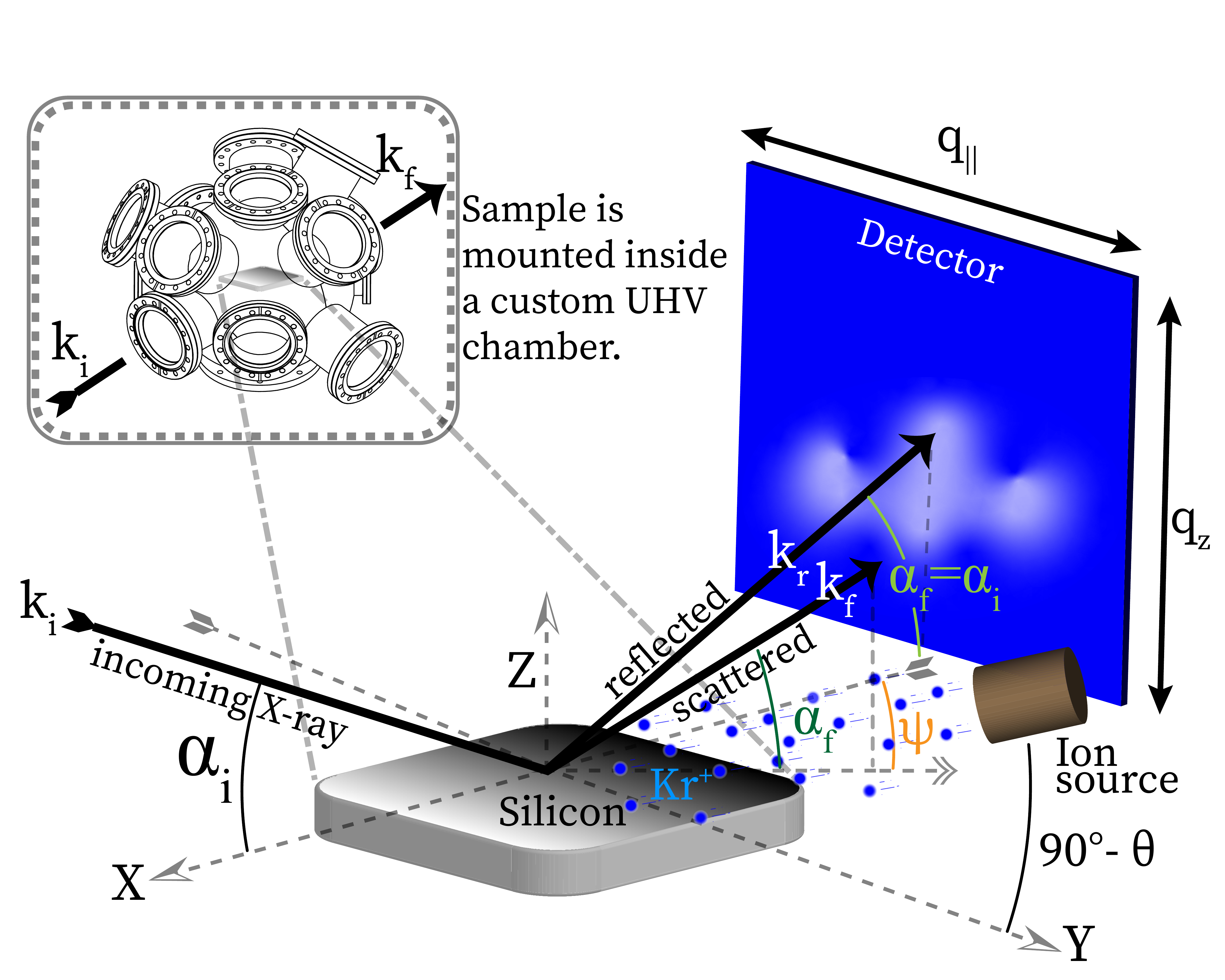}
\caption{\label{fig:GISAXS}  Schematic diagram of the GISAXS experiment. The ion source is placed at the polar angle $\theta$, which causes self-organized rippling on the silicon surface. The sample is positioned so that the X-ray incident angle $\alpha_i$ is slightly above the critical angle of total external reflection. The scattering is recorded as a function of the exit angles $\alpha_f$  and $\psi$ using a 2D detector.}
\end{figure}

As in Part I of this study, experiments utilized 640 $\mu$m thick p-doped (B) Si(100) wafers cut into 1 $\times$ 1  cm$^2$ pieces and cleaned with acetone, isopropyl alcohol, and methanol.  Samples were firmly affixed to a stage by Indium bonding. To prevent sputtering of impurities onto the surface, the sample stage geometry was designed to ensure that nothing was above the sample surface. The temperature of the water-cooled sample stage was monitored using a thermocouple and the stage was electrically isolated except for a wire leading out to an ammeter in order measure ion flux. The sample holder was mounted in a custom UHV chamber with mica X-ray windows and a base pressure of 5 $\times$ $10^{-7}$ Torr. Samples were kept at room temperature and bombarded with a broad beam of 1 keV Kr$^+$ ions generated by a 3-cm graphite-grid ion source from Veeco Instruments Inc., placed at 65$^{\circ}$ ion incidence angle ($\theta$), as indicated in Fig. \ref{fig:GISAXS}. This ion incidence angle was chosen because it is known to cause self-organized rippling on the silicon surface \cite{engler2014evolution}.  The ion beam flux was measured to be 1 $\times$ 10$^{15}$ ions cm$^{-2}$s$^{-1}$ at the operating chamber pressure of 1 $\times$ $10^{-4}$ Torr. The final fluence was 3.8 $\times$ 10$^{18}$ ions cm$^{-2}$.

Real-time coherent Grazing-Incidence Small-Angle X-ray Scattering (GISAXS) experiments were performed at the Coherent Hard X-ray (CHX) beamline at the National Synchrotron Light Source-II (NSLS-II) of Brookhaven National Laboratory. The photon energy of 9.65 keV (wavelength $\lambda = 0.1258 \; \mathrm{nm}$) was selected with a flux of approximately 5 $\times$ $10^{11}$ photon s$^{-1}$ and beam dimensions 10 $\times$ 10 $\mu$m$^2$. Experiments used an Eiger-X 4M detector (Dectris) with 75 $\mu$m pixel size, which was located 10.3 m from the sample. The incident X-ray angle $\alpha_i$ was 0.3$^{\circ}$, which is slightly above the critical angle of total external reflection for silicon of $\alpha_c =$ 0.186$^{\circ}$. The projected incident X-ray beam direction on the sample was perpendicular to the projected ion beam direction.  This allowed scattering in the GISAXS geometry to probe the dominant direction of ripple formation for the chosen ion bombardment angle ($\pm \; y$-direction in the coordinate system of Fig. \ref{fig:GISAXS}).  The diffuse scattering was recorded as a function of the exit angles $\alpha_f$  and $\psi$ using the 2D detector.  The change in X-ray wavevector $\mathbf{q}$ can be calculated from those angles:
\begin{equation}
\mathbf{q} = \mathbf{k_f}-\mathbf{k_i} = 
\begin{pmatrix}
  q_x  \\
  q_y  \\
  q_z 
 \end{pmatrix} = \frac{2\pi}{\lambda}
 \begin{pmatrix}
  \cos(\alpha_i) - \cos (\alpha_f) \cos(\psi)  \\
  \cos (\alpha_f) \sin (\psi)  \\
  \sin (\alpha_i)+ \sin (\alpha_f) 
 \end{pmatrix}
 \label{equ:wavenumber_conversion}
\end{equation} 
Since $q_x$ is small, the horizontal component $q_{||}$ (parallel to the surface) can be approximated as simply $q_y$ and the vertical component as $q_z$ (perpendicular to the surface). In the analysis of this paper, we will primarily be interested in the scattering along the Yoneda wing, which is particularly sensitive to surface structure \cite{renaud2009probing}. For simplicity, we use the term ``GISAXS pattern'' for the one-dimensional intensity curve $I(q_{||},t)$ obtained by averaging speckles along the Yoneda wing.

It is important to note that the coordinate system convention of Fig. \ref{fig:GISAXS} follows that often used for GISAXS experiments \textit{and is therefore rotated 90$^{\circ}$ with respect to the coordinate system typically used in the ion bombardment literature}.  Thus, in these experiments "parallel-mode" ripples form with their wavevector pointing in the y-direction rather than in the x-direction, as would conventionally be the situation in studies of ion beam nanopatterning.

\begin{figure}
\includegraphics[width=3.2 in]{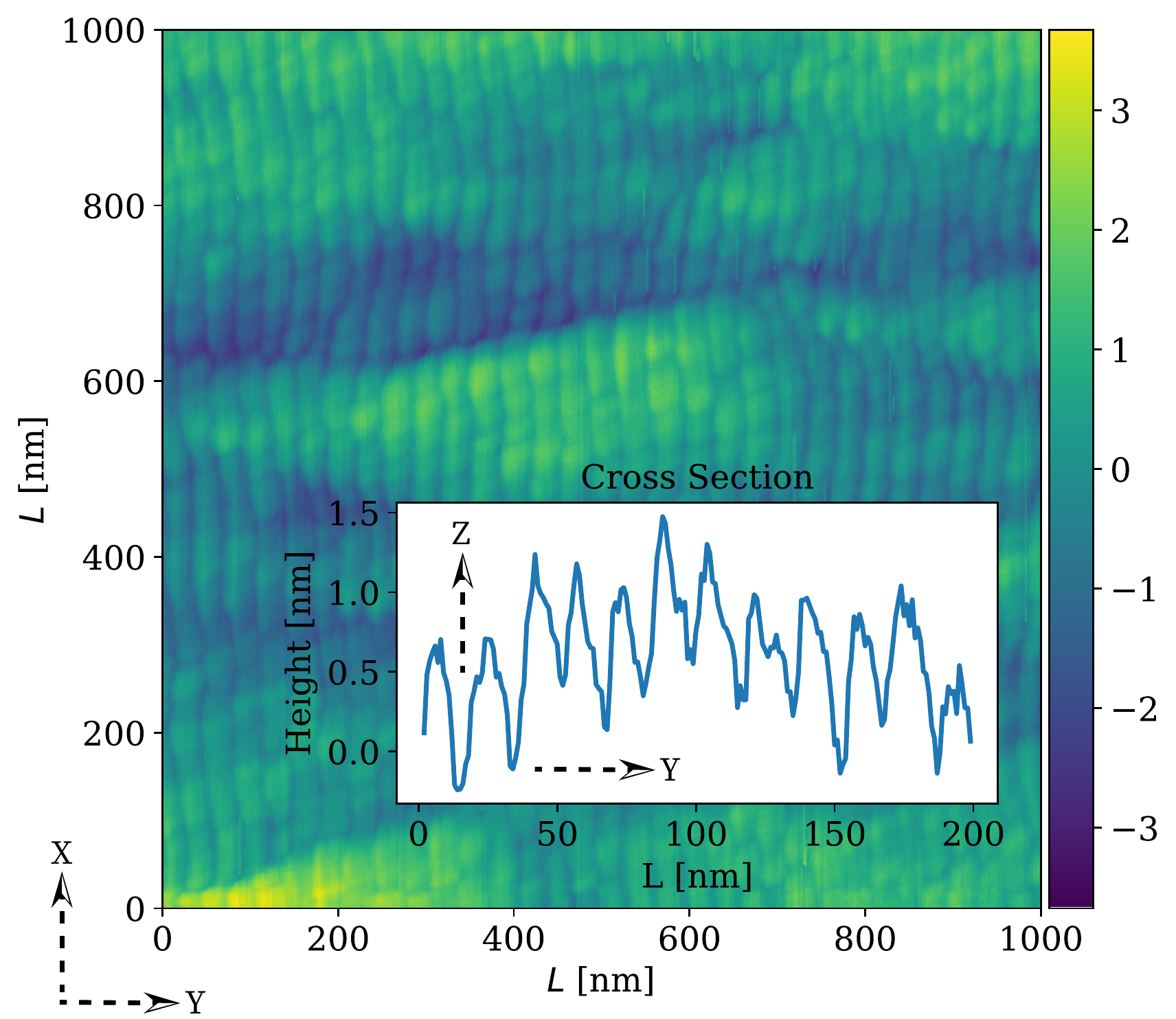}%

\includegraphics[width=2.8 in]{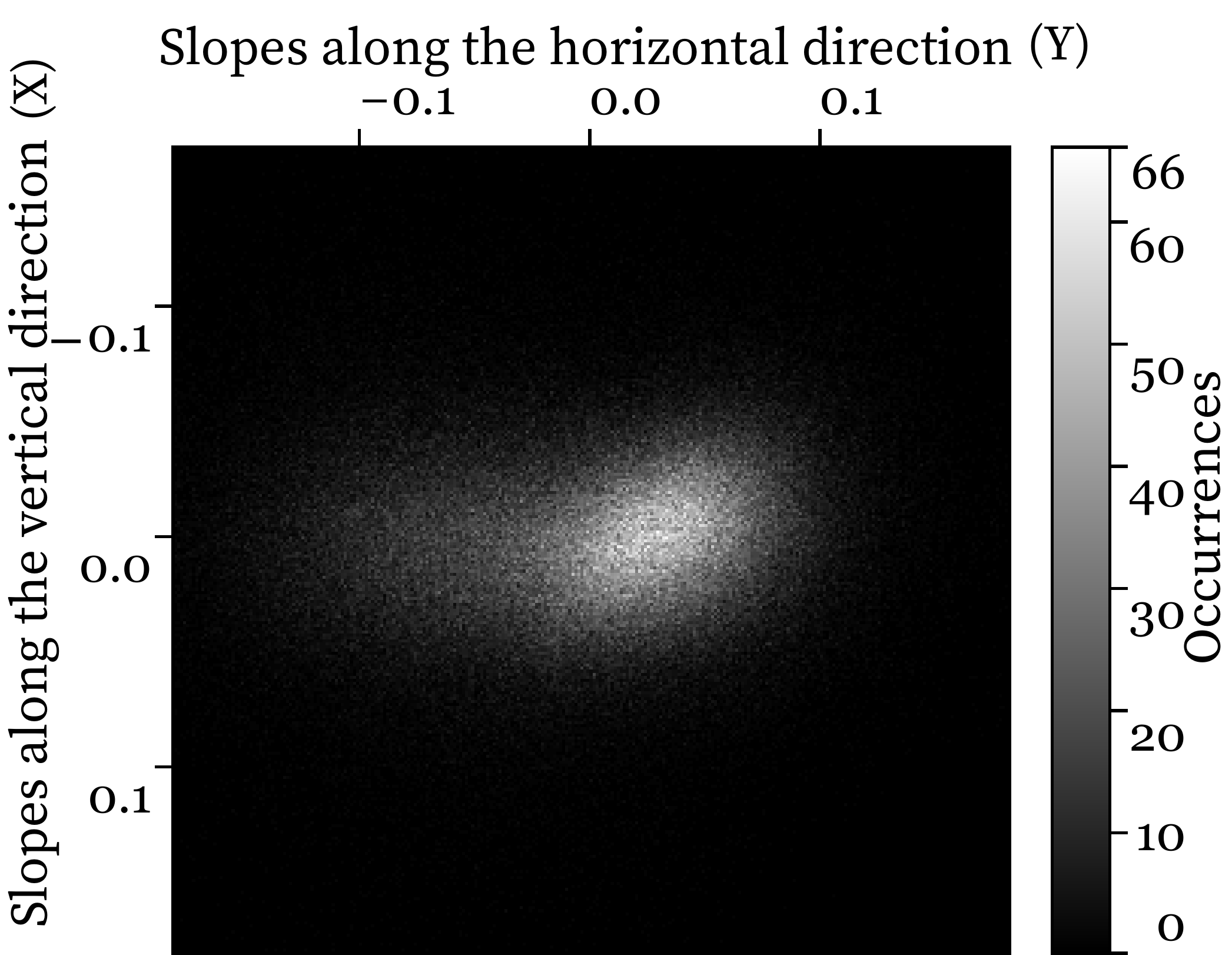}%
\caption{\label{fig:AFM}Top: \textit{Post facto} AFM image of silicon surface. The direction of the projection of ion beam's path onto the images is from right to left, while that of the X-ray path is from bottom to top. Bottom: The slope distribution calculated from the AFM image above; a denser distribution of positive slopes indicates that the slopes on the left side of the terraces, shown in the cross-section images, are more defined than the slopes on the right side.  The fluence was 3.8 $\times$ 10$^{18}$ ions cm$^{-2}$.}
\end{figure}

In order to examine the real space structure, \textit{post facto} AFM topographs were made of the sample; these show the development of nanoscale ripples (Fig. \ref{fig:AFM}).  In order to quantitatively compare the topographs with the X-ray experiments, the predicted GISAXS scattering patterns from the topographs were calculated based on an equation of Sinha \textit{et al.} \cite{sinha1988x}, as described in the present work's companion paper (Part I \cite{myint12020nanoscale}). This approach utilizes the z-component of the wave-vector change inside the material, calculated using the refracted incident $\alpha_i' = \sqrt{\alpha_i^2 - \alpha_c^2}$ and exit $\alpha_f' = \sqrt{\alpha_f^2 - \alpha_c^2}$ angles.  In this paper the geometrical value $q_z$ is used for display purposes in detector images since it is zero at the direct beam position on the detector, but in the data analysis, we use $q^\prime_z = 0.171 \; \mathrm{nm}^{-1}$ which is the average of $q_z^\prime$ along the Yoneda wing used in the analysis of the X-ray data.

\section{\label{sec:overview}Overview}

\begin{figure*}
\includegraphics[width=3.2 in]{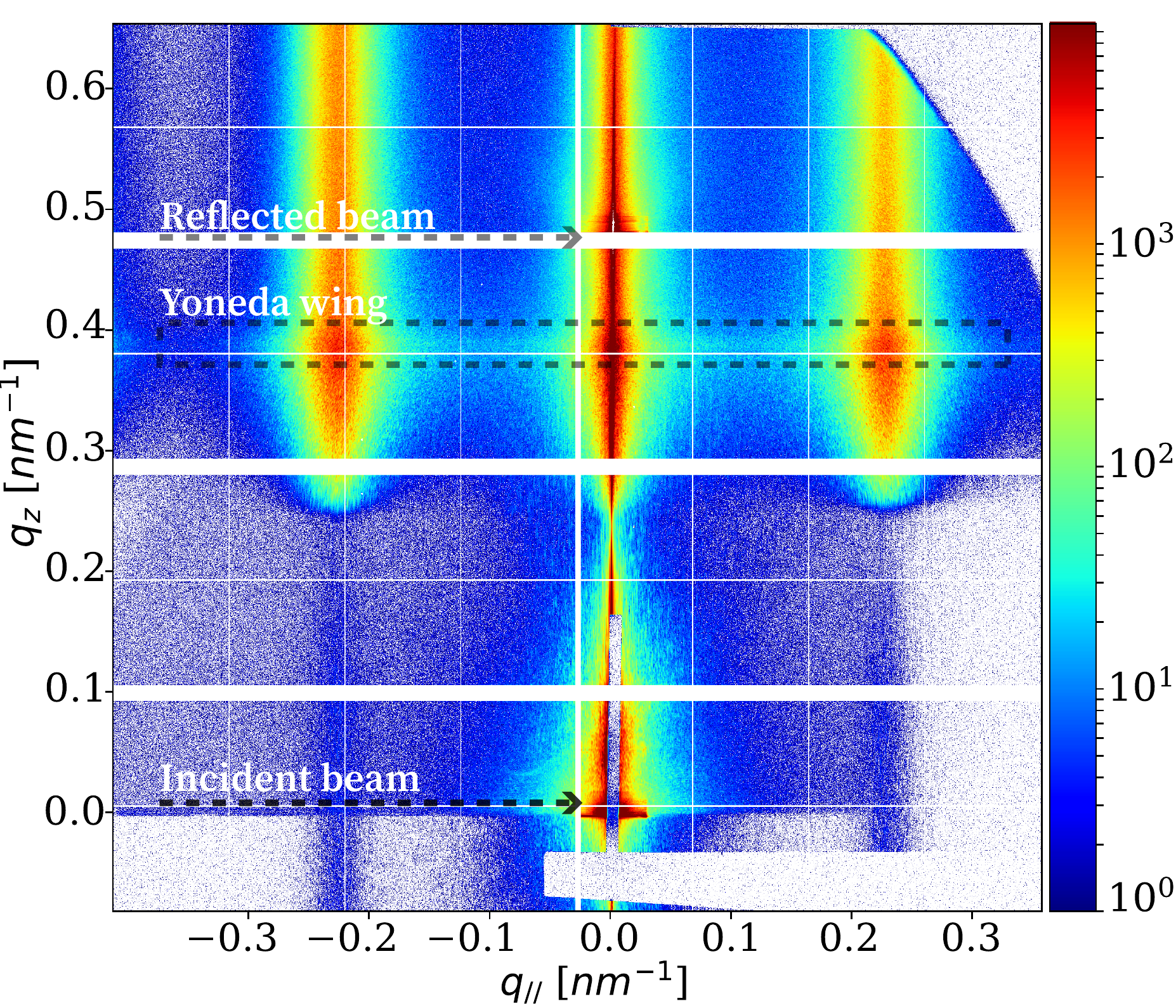}%
\includegraphics[width=3.2 in]{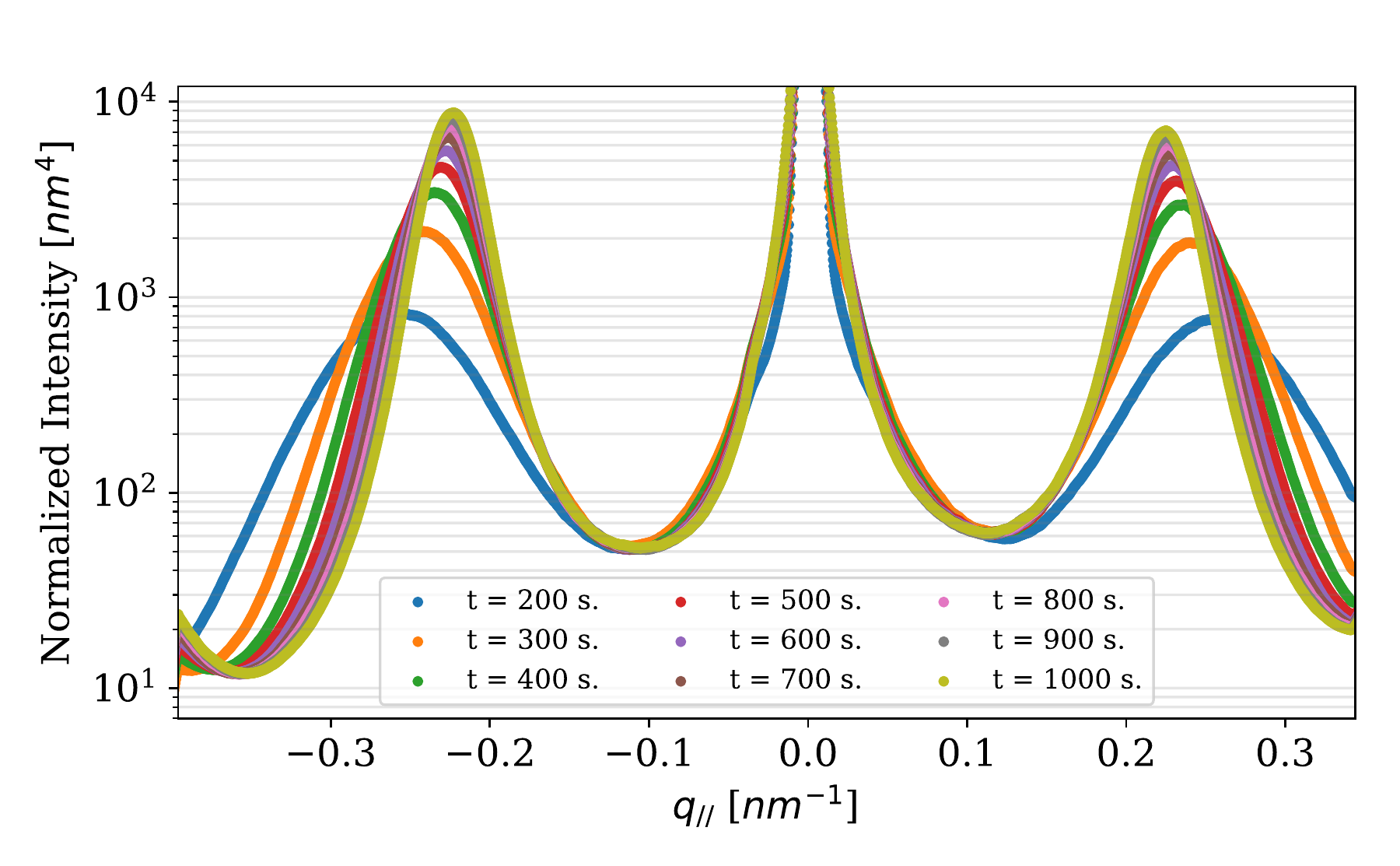}%
\caption{\label{fig:sq_Kr} Left: A typical detector scattering image during the nanopatterning. The Yoneda wing, spread across $q_{z}\, \simeq \,0.39\, \mathrm{nm}^{-1}$, is the highly surface-sensitive scattering exiting the sample at the critical angle, $\alpha_c$. Correlation peaks at $q_{||}\,\simeq\pm\,0.24\, \mathrm{nm}^{-1}$ are due to the formation of correlated nanoripples on the surface. Right: Evolution of GISAXS pattern obtained by averaging across the Yoneda wing, as indicated by the dotted box in the left diagram.}
\end{figure*}

In the experiments, ion bombardment started at $t$ = 0 s (after 100 s of static scan), and a clear correlation peak can be seen growing around $t = 200 \; \mathrm{s}$ due to the formation of correlated ripples on the surface.  The initial peak wavenumber is at $q_0 \approx \pm 0.26 \; \mathrm{nm}^{-1}$ so that the initial ripple wavelength is approximately  $2\pi/$0.26  nm$^{-1}$ $\approx$ 24 nm. In Sect. \ref{sec:early_kinetics} below, we quantitatively analyze this behavior using linear theory of nanopatterning. A typical detector pattern and GISAXS intensity patterns at particular times in the evolution are shown in Fig. \ref{fig:sq_Kr}.  After the initial ripple formation, coarsening occurs with the correlation peak position $\pm q_0$ shifting to a smaller wave number. The coarsening proceeds at an ever decreasing rate and, by the end of the experiment, the average GISAXS pattern changes only slowly; the final ripple wavelength calculated from the correlation peak position was approximately $2\pi/$0.23 nm$^{-1}$ $\approx$ 27 nm.

\section{\label{sec:early_kinetics}SPECKLE-AVERAGED EARLY-TIME KINETICS}
\begin{figure}
\includegraphics[clip,trim={0 0 0 0},width=3.41in]{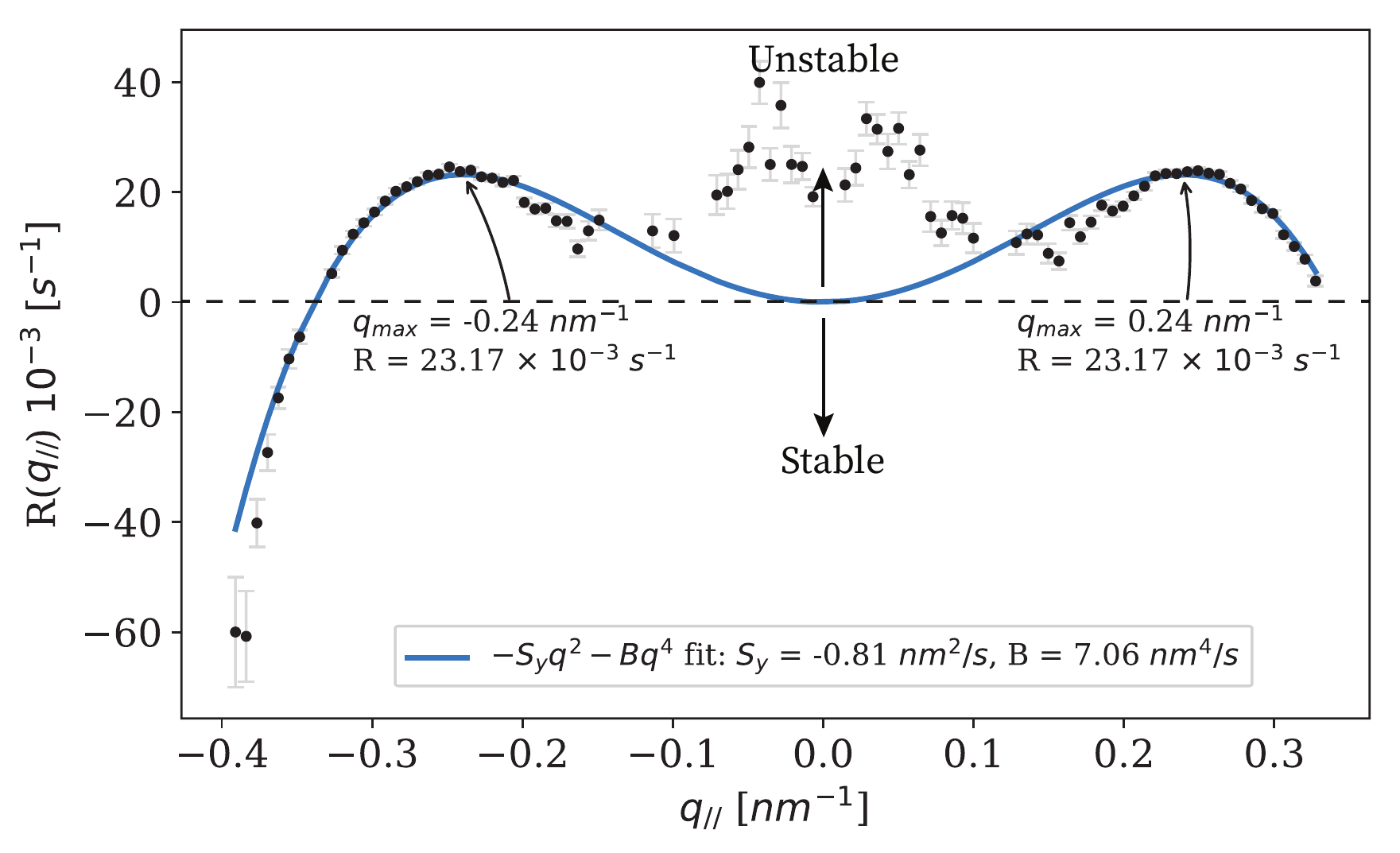}%

\includegraphics[width=3.41in]{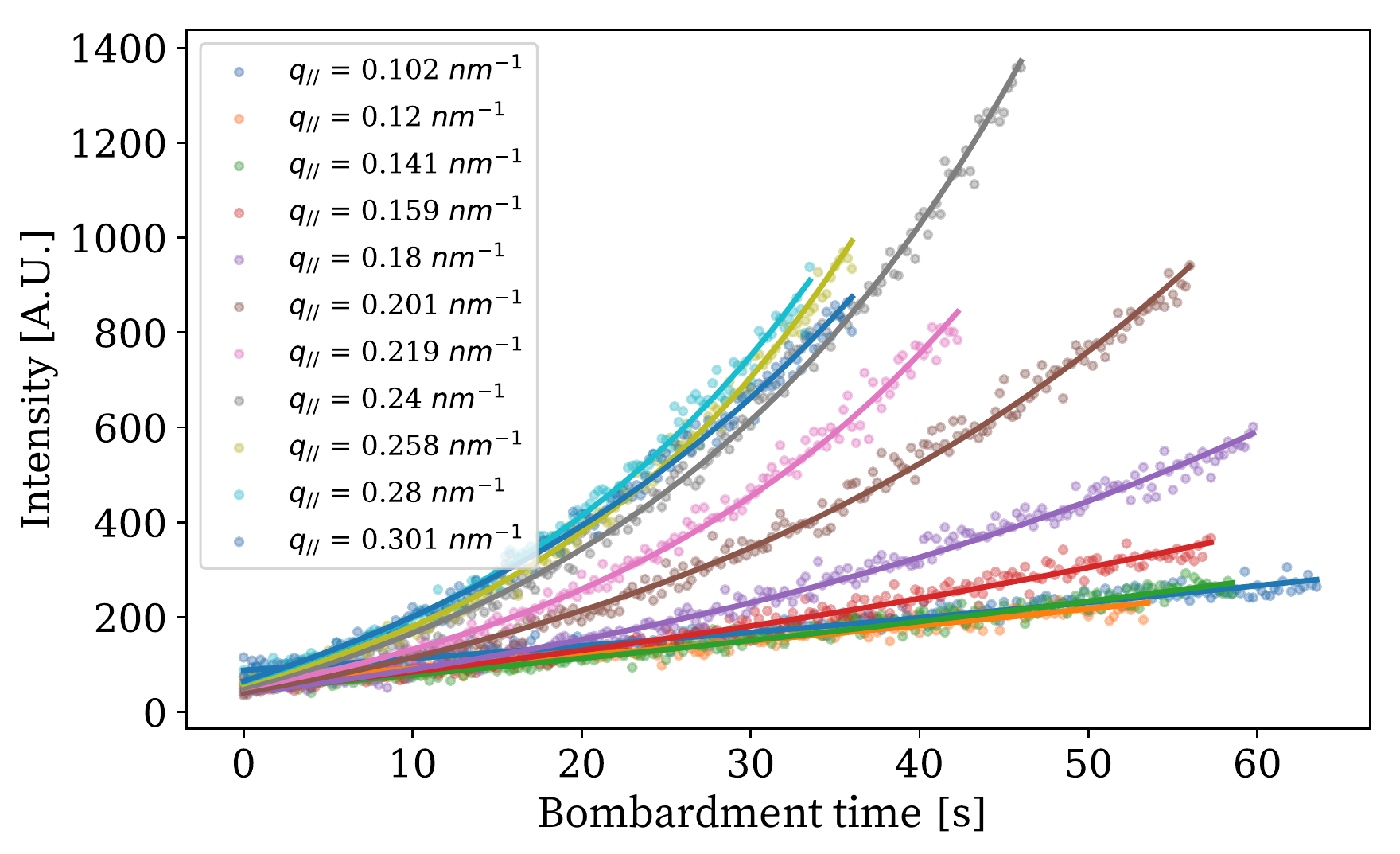}
\caption{\label{fig:linear-th-fits} Bottom: Simple exponential linear theory fits of intensity $I(t,q_{||})$ for different $q_{||}$.  Fitting cut-off times were determined by the criterion that the reduced $\chi^2$ of the fit be approximately one.  Top: Amplification factors $R(q_{||})$ extracted using the linear theory analysis with fit using Eq. \ref{equ:long-wave}.}
\end{figure}

As discussed in Part I, at early stages of nanopatterning, when surface slopes are small, a linear stability analysis can be applied to determine some of the coefficients in theories attempting to describe the process.  This linear theory applied to surface evolution during ion bombardment takes the form \cite{bradley1988theory}:
\begin{equation}
\frac{\partial\tilde{h}\left(\mathbf{q},t\right)}{\partial t}=R\left(\mathbf{q}\right)\tilde{h}\left(\mathbf{q},t\right)+\tilde{\eta}\left(\mathbf{q},t\right)
\label{eq: dispersion-general-form}
\end{equation}
where $\tilde{h}\left(\mathbf{q},t\right)$ is the Fourier transform of the surface height $h\left(\mathbf{r},t\right)$, $R\left(\mathbf{q}\right)$ is  the \emph{amplification factor}, and $\tilde{\eta}\left(\mathbf{q},t\right)$ is the Fourier transform of a stochastic noise.  The amplification factor can be determined experimentally by measuring the speckle-averaged height-height structure factor evolution \cite{madi2011mass,norris2017distinguishing}:

\begin{eqnarray}
\label{equ:hhstructure-factor}
I(\mathbf{q},t)&&= \left\langle h(\mathbf{q},t) \, h^*(\mathbf{q},t)\right\rangle\nonumber\\&& =\left(I_0(\mathbf{q})+\frac{n}{2R(\mathbf{q})}\right)e^{2R(\mathbf{q})t}-\frac{n}{2R(\mathbf{q})}
\end{eqnarray}
where $n$ is the magnitude of the stochastic noise: $\left\langle \eta\left(\mathbf{r},t\right) \eta\left(\mathbf{r^\prime},t\right) \right\rangle = n \, \delta(\mathbf{r}-\mathbf{r^\prime})\delta(t-t^\prime)$. The amplification factor differentiates surface stability or instability; a positive $R(\mathbf{q})$ at a given bombardment angle drives exponential amplification of modes of wavevector $\mathbf{q}$ resulting in surface instability, while a negative $R(\mathbf{q})$ damps fluctuations and stabilizes modes of wavevector $\mathbf{q}$. 

To determine $R(q_x \approx 0,q_{||}) \equiv R(q_{||})$, the intensity values $I(q_{||},t)$ at each wavenumber were averaged over 5 detector pixels in the $q_{||}$ direction and 100 pixels in the $q_z$ direction to remove speckle from the coherent scattering pattern.  The temporal evolution of each wavenumber bin was then fit with a function of the form $I(q_{||},t) = a(q_{||})  e^{2R(q_{||})t} + b(q_{||})$, with $a$, $b$ and $R$ being the fit parameters for each $q_{||}$ bin (Fig. \ref{fig:linear-th-fits}).

In the literature, the amplification factor is usually taken to have a form: 
\begin{equation}
R(q_{||})=-S_{y}\,q_{||}^2-B\,q_{||}^4
\label{equ:long-wave}
\end{equation}
where $S_y$ is a coefficient of curvature-dependent surface evolution and $B$ is a coefficient of ion induced surface viscous flow smoothening \cite{bradley1988theory,umbach2001spontaneous}. As seen in Fig. \ref{fig:linear-th-fits}, Eq. \ref{equ:long-wave} fits $R(q_{||})$ well within the range of wavenumbers measured. The bumps in $R(q_{||})$ at low $q_{||}$'s on each side of the GISAXS pattern are assumed to be due to overlap with tails of the specularly reflected X-ray beam and are not included in the $R(q_{||})$ fitting. Fit values are $S_y = -0.81 \; \mathrm{nm}^2 s^{-1}$ and $B = 7.1 \; \mathrm{nm}^4 s^{-1}$. The fastest growing wavenumber according to the linear theory should be $q^{max}_{||} = \sqrt{|S_y|/(2B)} = 0.24\; \mathrm{nm}^{-1}$, which is consistent with the early time peak position of the GISAXS profile. 

Comparisons between the fit parameter results found here for Kr$^+$ and those found earlier for Ar$^+$ patterning from Part I of this work are shown in Table \ref{tab:linear-th}.  The ion-enhanced viscous flow parameter $B$ is very similar, suggesting that the deposited energy is more important in determining viscous flow than is the momentum.  Moreover, as discussed in Part I, this magnitude of ion-enhanced viscous flow is consistent with previous measurements and with expectations if the effective viscosity is inversely proportional to ion flux, as discussed in Norris \textit {et al.} \cite{norris2017distinguishing}.

The similarity in the viscous relaxation parameters $B$ for Kr$^+$ and Ar$^+$ shows that the differing curvature coefficient $S_y$ between the two ions is the factor determining the difference in initial ripple length scales.  In particular, the magnitude of the (negative) curvature coefficient increases on going from Ar$^+$ to Kr$^+$.  

For theoretical comparison with the measured $S_y$, we examine the erosive formalism of Bradley and Harper \cite{bradley1988theory} and the redistributive formulism of Carter and Vishnyakov \cite{carter1996roughening}, while acknowledging that stress-driven theories offer competing views \cite{castro2012hydrodynamic,castro2012stress,norris2012stress,moreno2015nonuniversality,munoz2019stress}.  The erosive and redistributive contributions to $S_y$ can be estimated following the general approaches of Bobes \textit{et al.} \cite{bobes2012ion} and Hofs{\"a}ss \cite{hofsass2014simulation} using SDTrimSP \cite{mutzke2019sdtrimsp} binary collision approximation simulations. These give an erosive contribution $S_y^{eros} \approx 0.48$ nm$^2$/s and a redistributive contribution $S_y^{redist} \approx -1.83$ nm$^2$/s, for a total $S_y^{eros+redist} \approx -1.35$ nm$^2$/s. This is somewhat larger than the measured value of -0.81 nm$^2$/s, but estimates of the parameters entering theory must be considered approximate.  It's noteworthy, that these theoretical values of $S_y$ show the same trend as the experiment in going from the Ar$^+$ projectile (discussed in Part I of this work) to Kr$^+$.  The calculations suggest that the key difference in the $S_y$ values comes from the increased lateral mass redistribution for the higher mass/higher momentum Kr$^+$ (with $S_y^{redist} \approx -1.83$ nm$^2$/s) ion as compared to Ar$^+$ ($S_y^{redist} \approx -1.39$ nm$^2$/s).  Separately, a different approach to calculating the curvature coefficient \cite{norris2014pycraters} using the PyCraters Python framework \cite{PyCraters2017} for crater function analysis on the SDTrimSP results gives $S_y^{total} \approx -0.73$ nm$^2$/s, much closer to our measured value.

\begin{figure}
\includegraphics[width=3.2 in]{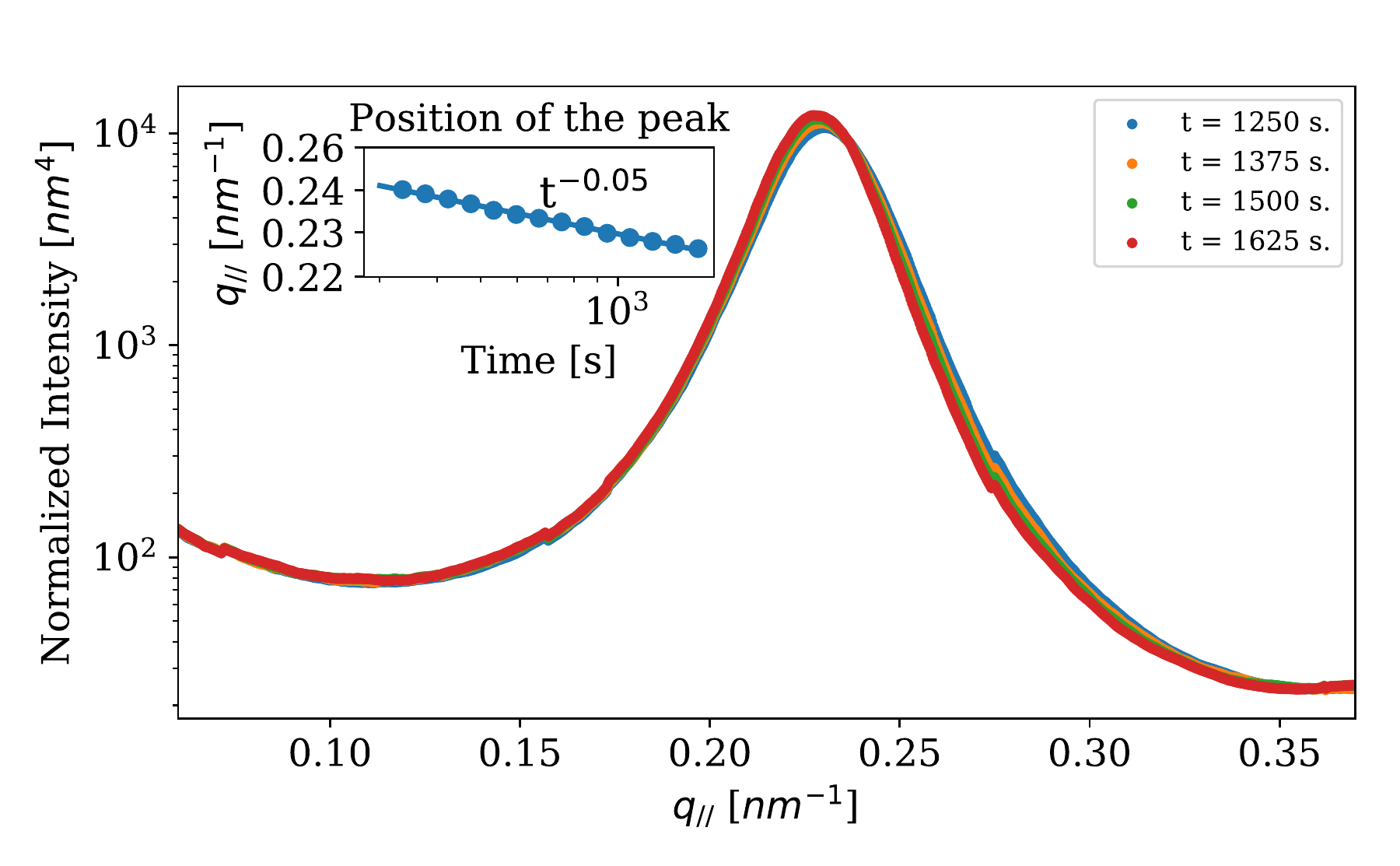}%
\caption{\label{fig:late_kinetics} Coarsening slows in the late stage of patterning. The inset shows a log-log plot of the correlation peak position $q_0$ as a function of time.}
\end{figure}

\begin{table}[b]
\caption{\label{tab:linear-th}%
Comparative linear theory analysis results}
\begin{ruledtabular}
\begin{tabular}{cccc}
Ion species &
\multicolumn{1}{c}{\textrm{$S_y$}}&
\multicolumn{1}{c}{\textrm{$B$}}&
\multicolumn{1}{c}{\textrm{$q_{max}$}}\\
 & [nm$^2s^{-1}$] & [nm$^4s^{-1}$] & [nm$^{-1}$] \\
\hline

Ar$^+$&-0.47&7.27&0.18  \\
Kr$^+$&-0.81& 7.06 & 0.24 \\


\end{tabular}
\end{ruledtabular}
\end{table}

\section{Speckle-Averaged Late-time kinetics and \textit{Post Facto} AFM}

With time, the ripple correlation peaks shift inward with an ever decreasing rate, showing that the spatial structure coarsens.  Beyond $t$ = 1000 s, the GISAXS pattern changes very little - the peak moves only a few pixels as shown in Fig. \ref{fig:late_kinetics}. While the range of time scales available is too limited to make a definitive statement about the nature of the relaxation, the peak motion can be fit as a weak power law evolution.
At late times, it's well known that the ripples begin to form asymmetric sawtooth structures.  As a result, the scattering pattern becomes asymmetric \cite{ludwig2002si,perkinson2018sawtooth}.  Here it's observed in Fig. \ref{fig:sq_Kr} that the correlation peak at $+ q_0$ grows slightly higher than the one at $ - q_0$.  More insight comes from the \textit{post facto} AFM topograph, which shows the asymmetric structure, as evidenced by the cut through the topograph and the slope analysis shown inthe slope analysis shown in Fig. \ref{fig:AFM}.  The slope analysis also shows that the positive slopes are more uniform in angle distribution than are the negative slopes.  

Simple calculations of the scattering expected from a sawtooth structure show that, if the positive terrace slope is larger in magnitude than the negative terrace slope on the structure, the positive $q_{||}$ peak should be higher, as observed. In this case, the negative terrace slope is facing the incoming ion beam.  Such calculations also show that harmonic peaks should exist at $\pm 2q_0$, with the peak at $-2q_0$ higher than the one at $+2q_0$.  While the harmonic peaks are off the detector, the increase in intensity on each side of the detector is leading up to them, with the increase higher on the negative side as expected from the calculations.

\begin{figure}
\includegraphics[width=3.2 in]{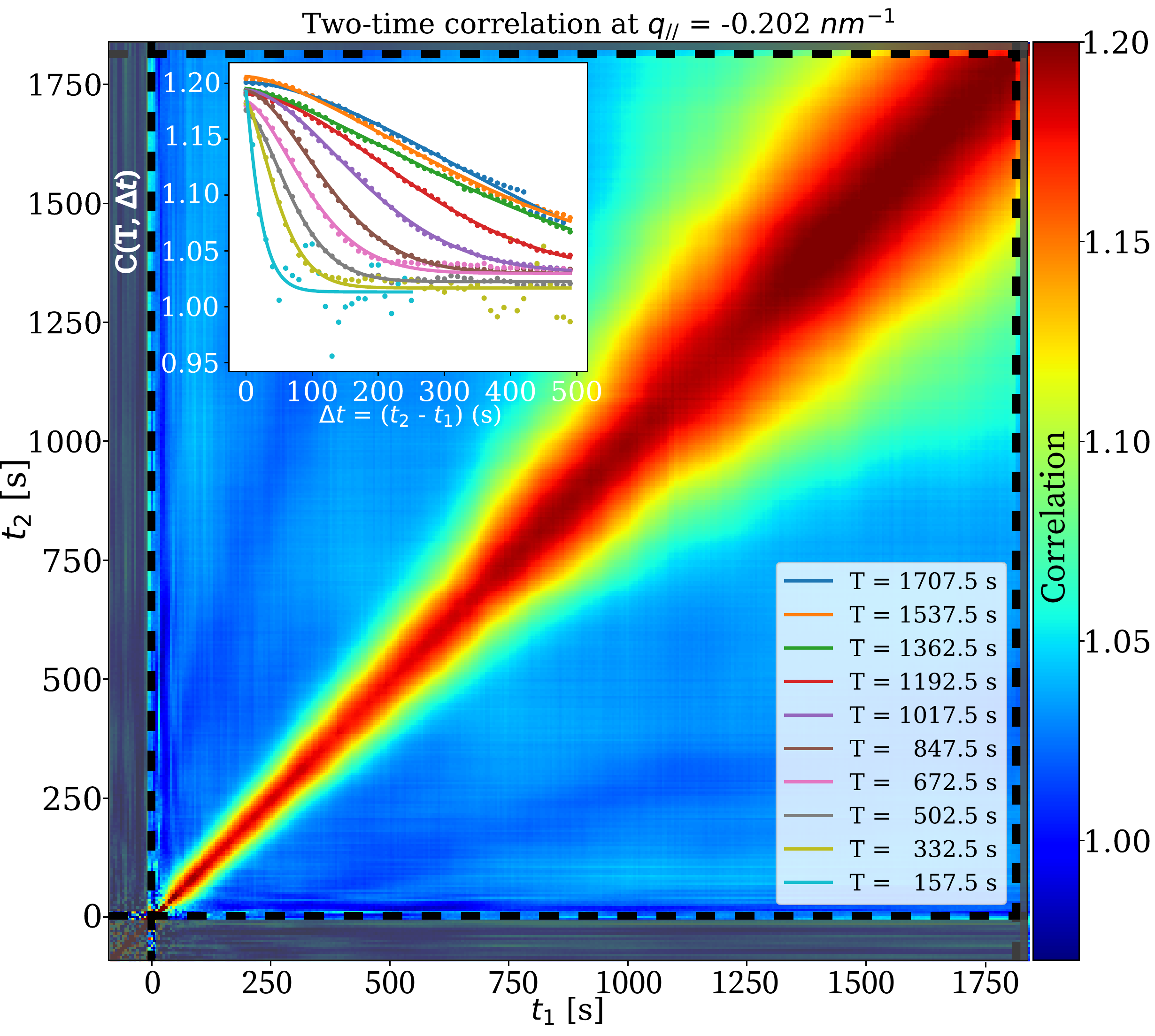}
\caption{\label{fig:TT} Evolution of a two-time correlation function (TTCF) at a wavenumber near the scattering peak $q_0$. The surface was originally smooth with little scattering; the ion bombardment started at t = 0 s (after 100 s of static scan indicated by gray area).  The inset shows KWW fits to diagonal cuts of the TTCF as discussed in the text.}
\end{figure}

\begin{figure*}
\includegraphics[clip,trim={0.0in 0 0 0},width=3.41in]{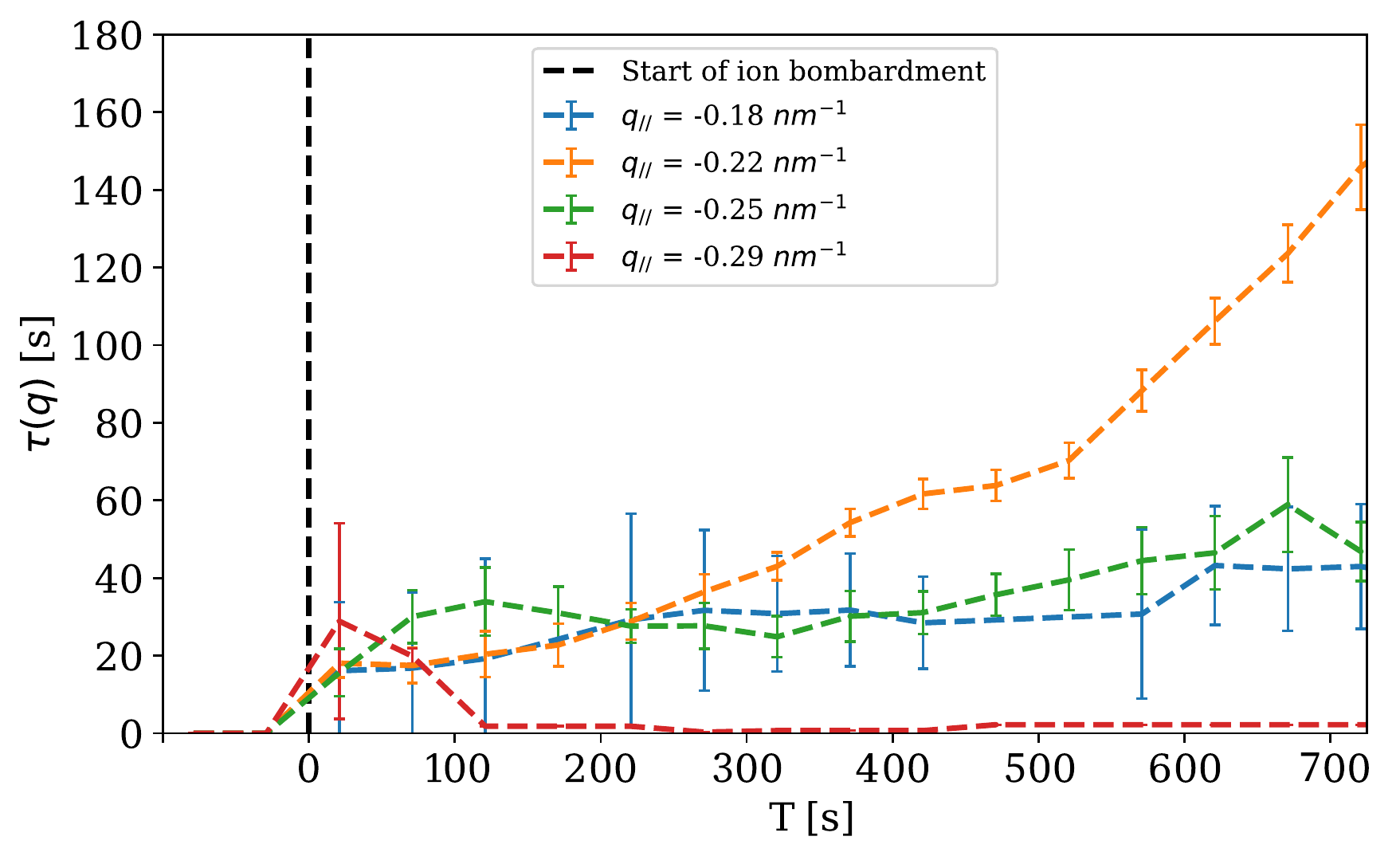}
\includegraphics[clip,trim={0.0in 0 0 0},width=3.41in]{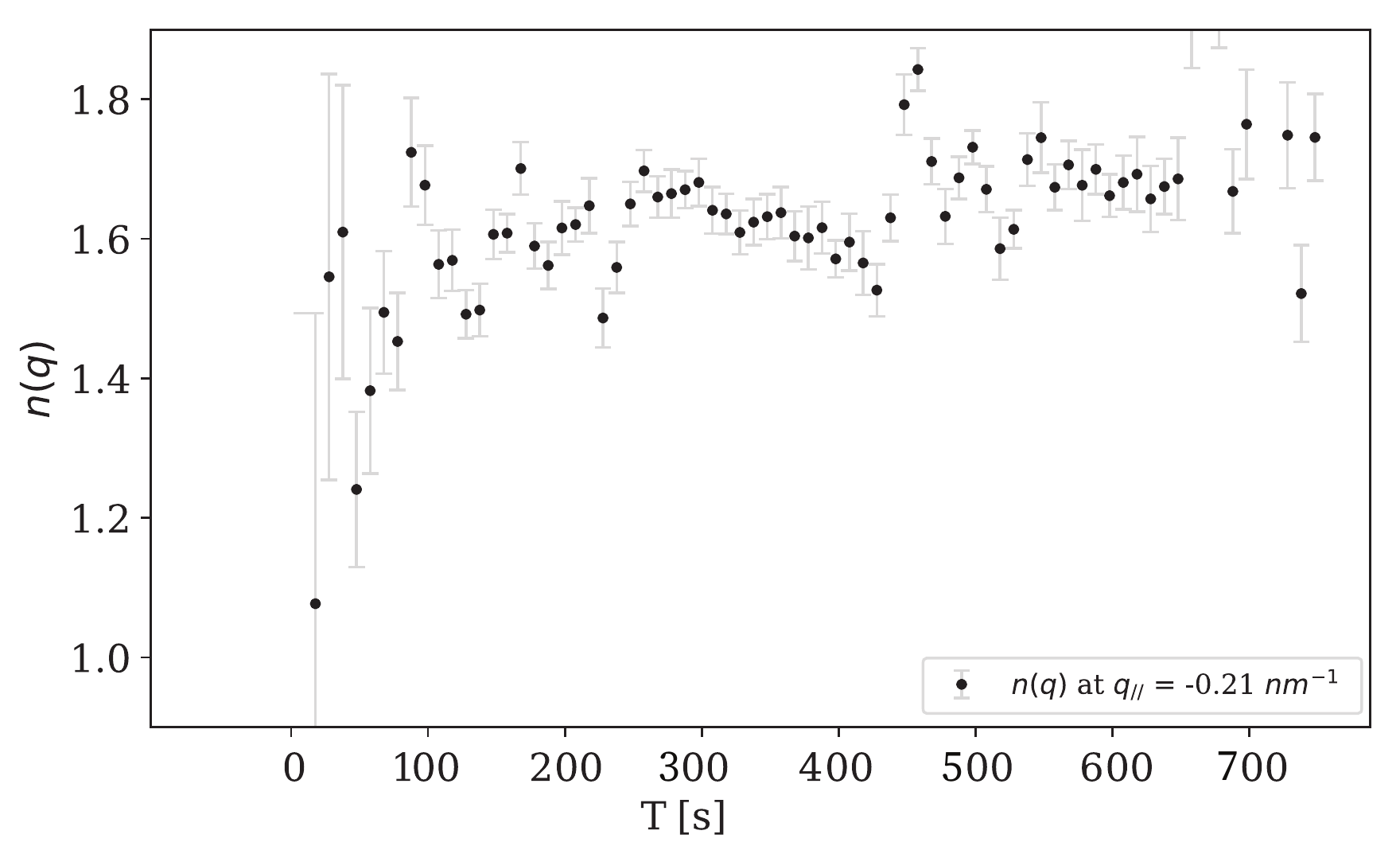}

\caption{\label{fig:TT_slice_selectq} KWW equation fit results for diagonal slices through the TTCF's during nanopatterning. During the time period shown, the peak position decreases from $q_0 \; \approx$ 0.25 nm$^{=-1}$ to $q_0 \; \approx$ 0.22 nm$^{-1}$. Left: Evolution of correlation time $\tau(q_{||})$.  Right: Evolution of exponent $n(q_{||})$ at wavenumber $q_0$ = 0.21 nm$^{-1}$. Different adjacent time averaging was performed here to highlight the quick transition of $n(q_{||})$ from 1 to a value of 1.6-1.8.}
\end{figure*}

\begin{figure*}
\includegraphics[clip,trim={0.0in 0 0 0},width=7.0in]{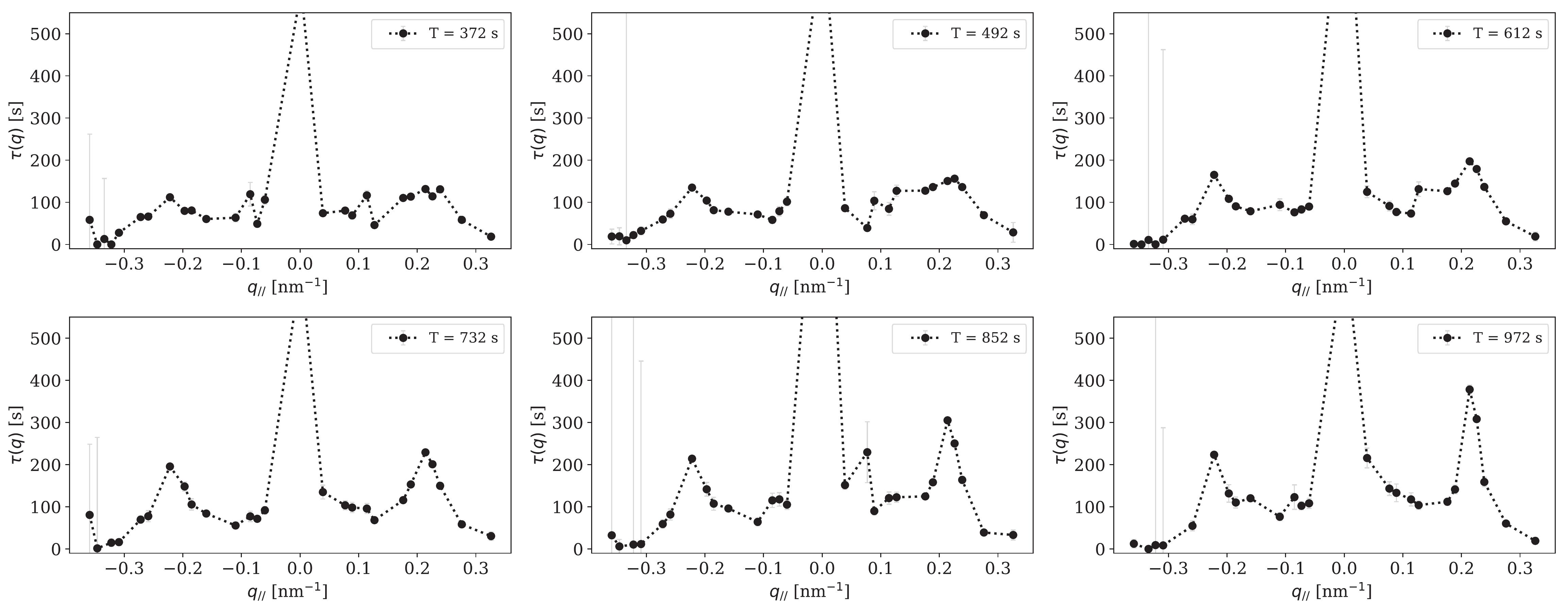}%
\caption{\label{fig:TT_slice_allq_Kr} Plots of $\tau(q_{||})$ at different bombardment times $T$ obtained from the KWW fits of TTCF diagonal cuts.  Near the origin, the correlation times can be unreliable due to overlap between the scattered signal and the tail of the specular reflection.  The fit value of $\tau$ at the origin is quite high, perhaps as a result of this.}
\end{figure*}

\section{Speckle Correlation Study of fluctuation dynamics}

Although the speckle-averaged GISAXS pattern shows the average kinetics, the strength of coherent experiments lies in their ability to measure temporal correlations of the  detailed speckle pattern, illuminating the underlying fluctuation dynamics. The two-time correlation function (TTCF) measures how the structure on a given length scale changes between time $t_1$ and time $t_2$ as the sample evolves:
\begin{equation}
C(q_{||},t_1,t_2)= \frac{\left\langle I(q_{||},t_1)I(q_{||},t_2)\right\rangle }{\left\langle I(q_{||},t_1)\right\rangle \left\langle I(q_{||},t_2)\right\rangle} 
\label{equ:twotime}
\end{equation}
where the angular brackets denote an average over equivalent $q_{||}$ values and the denominator values can be considered as speckle-averaged intensities one would have obtained using non-coherent scattering. 

TTCF's are shown in Fig. \ref{fig:TT} for a wavenumber $q_{||}$ near the scattering peak. The central diagonal ridge of correlation going from the bottom left to top right indicates the high correlation expected for $t_1 \approx t_2$. One way to understand how a surface changes on a given length scale is by observing the width of the central correlation ridge, which is a measure of correlation time on the surface. As seen in Fig. \ref{fig:TT}, near $q_0$, the width increases continuously during the experiment. However, as discussed below, at wavenumbers away from $q_0$, the TTCF width initially increases rapidly but gradually approaches an approximate steady state.

Quantitative measurement of the evolving dynamics is made by taking diagonal cuts through the central ridge at a constant average bombardment time $T = (t_1+t_2)/2$ as a function of $\Delta t = |t_2 - t_2|$ at each wavenumber $q_{||}$. The decay in correlation with time is fit with the Kohlrausch-Williams-Watts (KWW) form\cite{williams1970non}:
\begin{equation}
g_2^T(q_{||},\Delta t)= b+\beta \, e^{-2({\frac{\Delta t}{\tau(q_{||})}})^{n(q_{||})} },
\label{equ:KWW}
\end{equation}
where $\tau(q_{||})$ is the fit correlation time, and $n(q_{||})$ is a fit exponent which shows whether the relaxation is a simple ($n$ = 1), stretched ($0 < n < 1$), or compressed ($n > 1$) exponential. $b$ is the baseline, which was set as 1 or allowed to vary between 0.9 - 1.1. $\beta(q_{||})$ describes the contrast, which depends on experimental factors including the effective resolution of the experiment.  The magnitude of the central diagonal ridge of correlations in Fig. \ref{fig:TT} increases with time, indicating an increasing contrast.  This is probably because background incoherent scattering (e.g. from slits or windows) causes the apparent contrast to decrease at early times when the scattering from the sample is relatively small.  As ripples form on the sample, the scattering from the sample increases and the apparent contrast approaches its limiting value.  Finally, to improve statistics for the fits to Eq. \ref{equ:KWW}, results from $\pm$ 10 s around the central mean growth time $T$ were chosen for averaging.

Figure \ref{fig:TT_slice_selectq} shows the evolution of $\tau(q_{||})$ and $n(q_{||})$ for selected wavenumbers.  Near the peak wavenumber $q_0$, $\tau(q_{||})$ increases continuously.  The exponent $n(q_{||})$ rapidly increases from approximately one, indicative of simple exponential decay, to a value of 1.6-1.8, showing compressed exponential behavior.  Away from $q_0$, the $\tau(q_{||})$ values initially increase but then seem to relax to a steady state.

Figure \ref{fig:TT_slice_allq_Kr} complements Fig. \ref{fig:TT_slice_selectq} by showing the behavior of $\tau(q_{||})$ for selected times.  It's seen that the $\tau(q_{||})$ values near the scattering peaks $\pm q_0$ grow strongly to become much larger than the relaxation times at other wavenumbers. Starting at $T$ = 852 s, it is observed that $\tau(q_{||})$ is asymmetric, being higher at $+q_0$ than at $-q_0$.

More detail can be obtained from averaging over larger time periods of $T = 200-450 \;\mathrm{s}, 650-900 \;\mathrm{s}, 1400-1650 \;\mathrm{s}$, i.e. mean $T = 325 \;\mathrm{s}, 775 \;\mathrm{s}, 1525 \;\mathrm{s}$, using the auto-correlation function:
\begin{equation}
g_2(q_{||},\Delta t)= \frac{\left\langle I(q_{||},t^{\prime})I(q_{||},t^{\prime}+\Delta t)\right\rangle }{\left\langle I(q_{||}) \right\rangle ^2}.
\label{equ:g2}
\end{equation}
The angular brackets indicate a time averaging over $t^\prime$ and equivalent $q$ values. The $g_2(q_{||},\Delta t)$ functions were fit with the KWW form Eq. \ref{equ:KWW} to obtain $\tau(q_{||})$ and $n(q_{||})$ values. 

\begin{figure*}
\includegraphics[width=3.1 in]{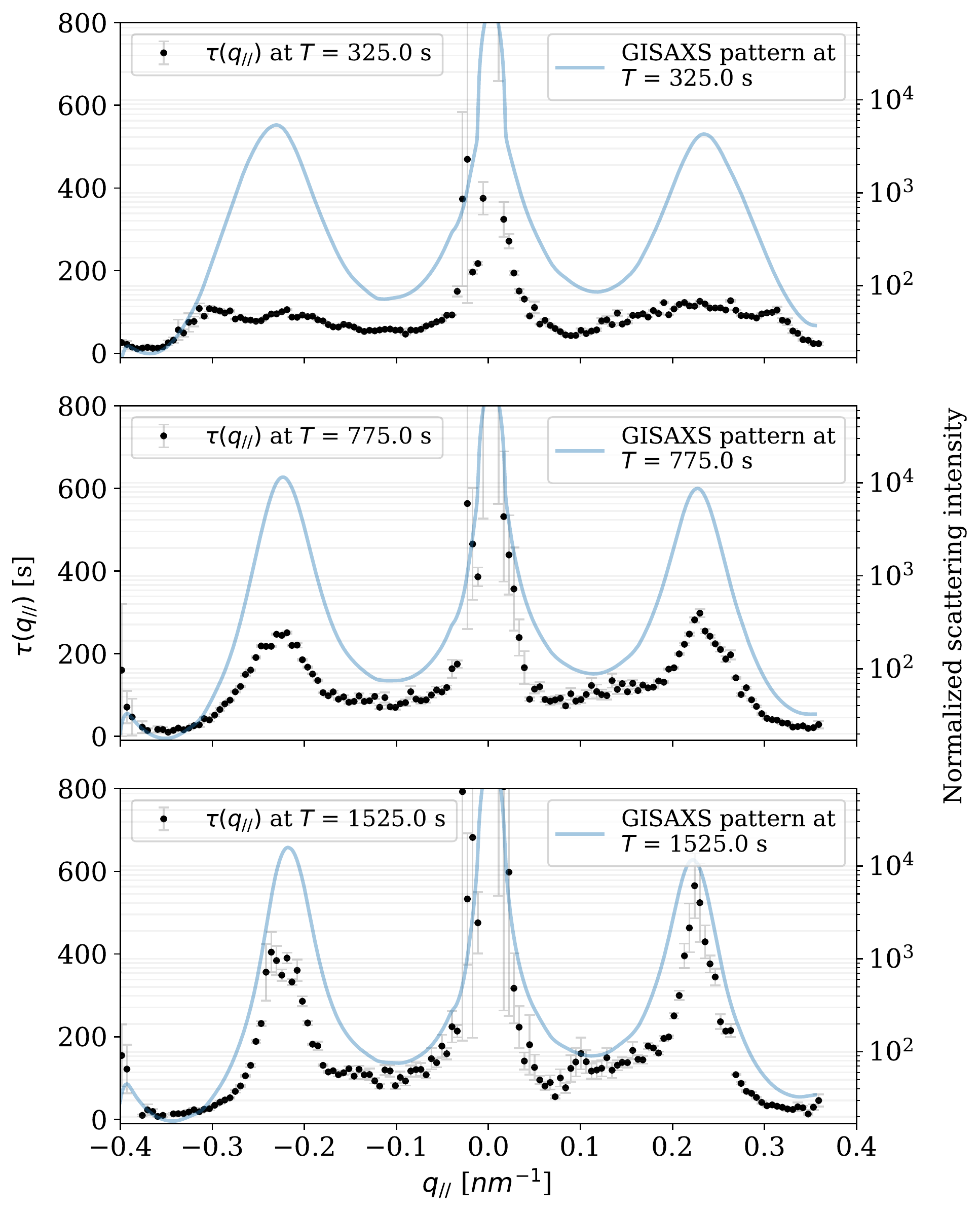}%
\hspace{0.05in}
\includegraphics[width=3.1 in]{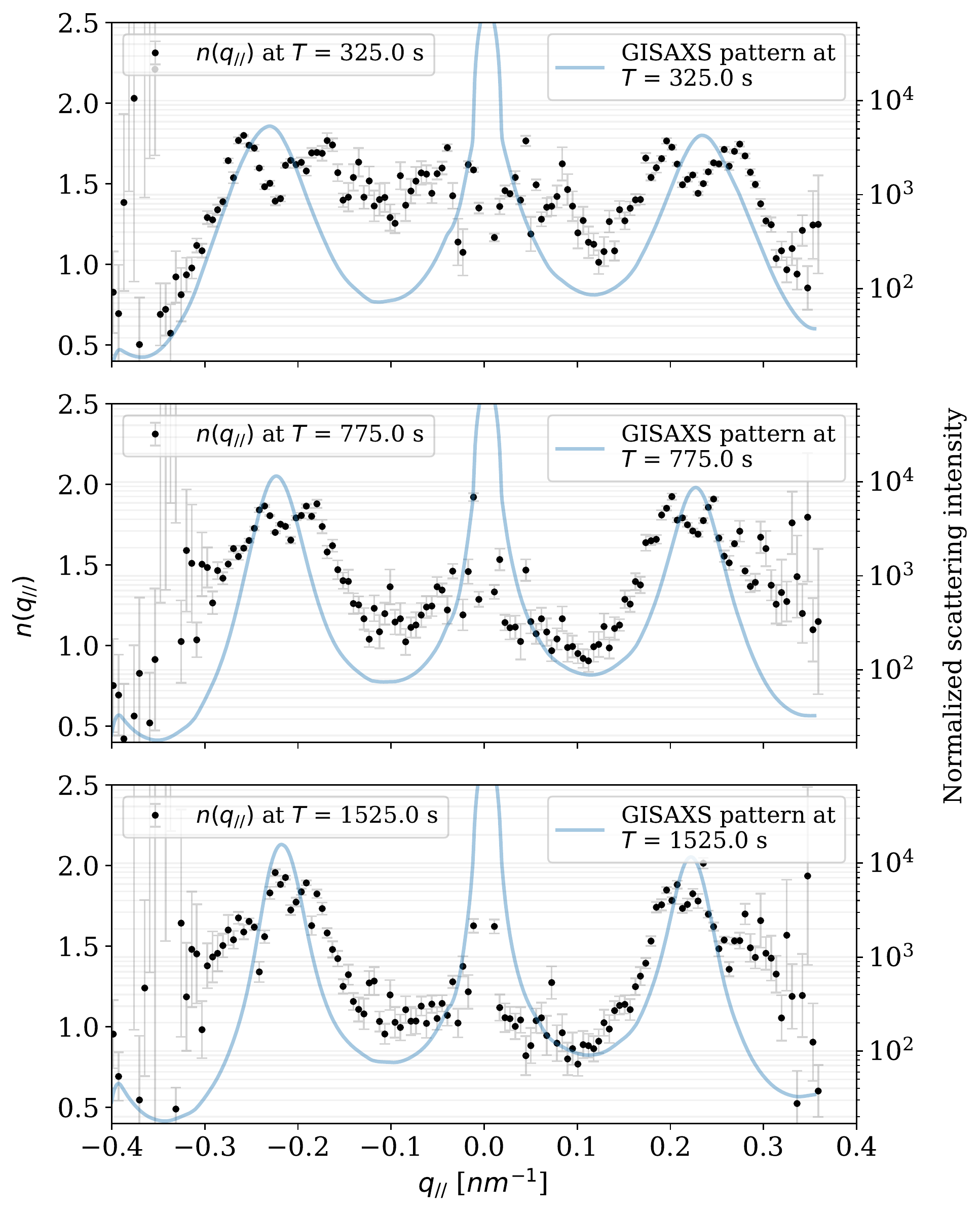}%
\caption{\label{fig:Kr_g2} Fit values of correlation time $\tau(q_{||})$ (left) and exponent $n(q_{||})$ (right) for selected bombardment times $T$ with larger time averaging and use of the autocorrelation function in Eq. \ref{equ:g2}.  For comparison, the GISAXS pattern is shown on a logarithmic intensity scale.}
\end{figure*}
Plots of experimental $\tau(q_{||})$ and $n(q_{||})$ are shown in Fig. \ref{fig:Kr_g2}. The trends seen in Fig. \ref{fig:TT_slice_allq_Kr} are confirmed and amplified.  Now it is more evident that the correlation time $\tau(q_{||})$ is asymmetric. The asymmetry is in the the opposite direction of the relative peak intensities.  In addition, from the upturn in $\tau(q_{||})$ near the detector edges, it appears that there is also a peak in $\tau(q_{||})$ at the harmonic peaks $\pm 2q_0$. Near the primary peaks, $n(q_{||}) > 1$, so that the relaxation is compressed exponential, as noted before.  At higher values of $q_{||}$, $n(q_{||})$ decreases to below one, indicative of stretched exponential behavior.  All of these behaviors are similar to those observed in Part I of this work for Ar$^+$ nanopatterning of Si.

\section{\label{sec:ripvelocity}ripple velocity analysis}
Following their self-organized growth, ripples can move across the surface, driven by the continued ion bombardment \cite{alkemade2006propulsion}. For uniform motion of waves or a uniform flow pattern, homodyne X-ray scattering cannot detect the motion except perhaps due to small edge effects.  However, following the work of Lhermitte \textit{et al.} with fluid flow \cite{lhermitte2017velocity}, we have shown in Mokhtarzadeh \textit{et al.} \cite{mokhtarzadeh2019nanoscale} that the ripple flow can be measured as systematic speckle motion on the detector if the ripple velocity field is inhomogeneous.

For self-organized silicon nanopatterning at room temperature, surface evolution is driven by the ion beam, so the local ripple velocity is expected to be proportional to the local ion flux.  The contribution to the velocity of ripples due to angle-dependent sputter erosion is \cite{bradley1988theory,alkemade2006propulsion}:
\begin{equation}
v_y(\theta) = - F \, \Omega \, \left[ \cos\theta \frac{dY(\theta)}{d\theta}-Y(\theta) \sin\theta \right]
\label{equ:velocity}
\end{equation}
where $Y(\theta)$ is the angle-dependent sputter yield, $F$ is the ion flux and $\Omega$ is the atomic volume. The accuracy of Eq. \ref{equ:velocity} has been the subject of some controversy.  Given typical shapes of the sputter yield $Y(\theta)$, Eq. \ref{equ:velocity} predicts that there is a cross-over bombardment angle $\theta_c$, below which ripples move into the projected direction of the ion beam, and above which they move away from the ion beam. The exact value of $\theta_c$ depends sensitively on the detailed form of $Y(\theta)$. A number of FIB/SEM studies using Ga$^+$ ions have found ripple motion in the opposite direction to what is expected from the equation \cite{alkemade2006propulsion,habenicht2002ripple,wei2009propagation,gnaser2012propagation,kramczynski2014wavelength}. However, Hofs{\"a}ss \textit{et al.} \cite{hofsass2013propagation} measured the ripple velocity during 10 keV Xe$^+$ irradiation of Si using fabricated marker grooves with \textit{ex situ} SEM and found the predicted transition from ripple movement into the beam to the opposite direction with increasing bombardment angle.

If the ion flux $F$ is inhomogeneous, then Eq. \ref{equ:velocity} shows that the ripple velocity is as well.  For the present studies, a highly focused 10 x 10 $\mu\mathrm{m}^2$ X-ray beam was used.  However, in the coordinate system of Fig. \ref{fig:GISAXS}, the X-ray footprint on the sample is elongated by a factor of 200 in the $x$-direction by the grazing incidence geometry.  Therefore, given the small size of the x-ray beam, only the variation in ion intensity along the $x$-direction is important.  In earlier work \cite{mokhtarzadeh2019nanoscale} we had examined the case in which the ion beam is centered within the x-ray footprint.  That presented complexities however, which limited our ability to definitively analyze the results.  In the present experiment, the center of the x-ray beam is outside the x-ray footprint and we can successfully analyze the resulting data assuming a simple uniform gradient of ion flux in the $x$-direction, and thus a uniform gradient in ripple velocity $v_y(x)$ in the flow pattern sampled by the x-ray footprint. A second effect of a gradient in ion flux is an $x$-gradient in the erosion rate of the material $v_z(x)$. Note that, in the coordinate system of Fig. \ref{fig:GISAXS}, the erosion velocity is negative. It is related to the sputter yield by $v_z(x) = - F(x) \Omega Y(\theta) \cos\theta$.

In general, a uniform gradient in velocity can be written:
\begin{equation}
\mathbf{v}(r) = \mathbf{v}_0(r) + \mathbf{\Gamma} \cdotp \mathbf{r}
\end{equation}
with $\bf \Gamma$ being the velocity gradient tensor.  Fuller \textit{et al.} showed that, as a result of the flow pattern, speckles move in reciprocal space as \cite{fuller1980measurement}:
\begin{equation}
\frac{d\mathbf{q}}{dt} = -\mathbf{\Gamma^{\mathbf{T}}} \cdotp \mathbf{q}
\label{equ:dqdt}
\end{equation}
In the geometry of the present experiment, this gives:
\begin{equation}
\frac{dq_x}{dt}(t)= -\Gamma_y(t) q_{||} - \Gamma_z(t) q_z,
\label{equ:dqx}
\end{equation}
where we have explicitly included the possibility of a time-dependence to the speckle velocity in writing $dq_x/dt(t)$.  Here $\Gamma_y(t)=\Delta v_y(x,t)/\Delta x$ is the (time-dependent) $x$-gradient in the ripple velocity and $\Gamma_z(t)=\Delta v_z(x,t)/\Delta x$ is the (time-dependent) gradient in (negative) erosion velocity.

On the detector, speckle motion in the $x$-direction shows as vertical movement. Indeed, as discussed below, this behavior is observed in the data and supports the supposition that there is a simple unidirectional gradient of ion flux across the footprint of the X-ray beam in the $x$-direction.  This is also an important difference from our earlier work of Ref. \cite{mokhtarzadeh2019nanoscale} in which a highly focused ion beam was used so that the X-ray beam straddled both sides of the ion beam distribution.  As shown below, the present experimental arrangement allows us to determine both the direction and speed of the surface ripples.

\begin{figure}
\includegraphics[width=3.1 in]{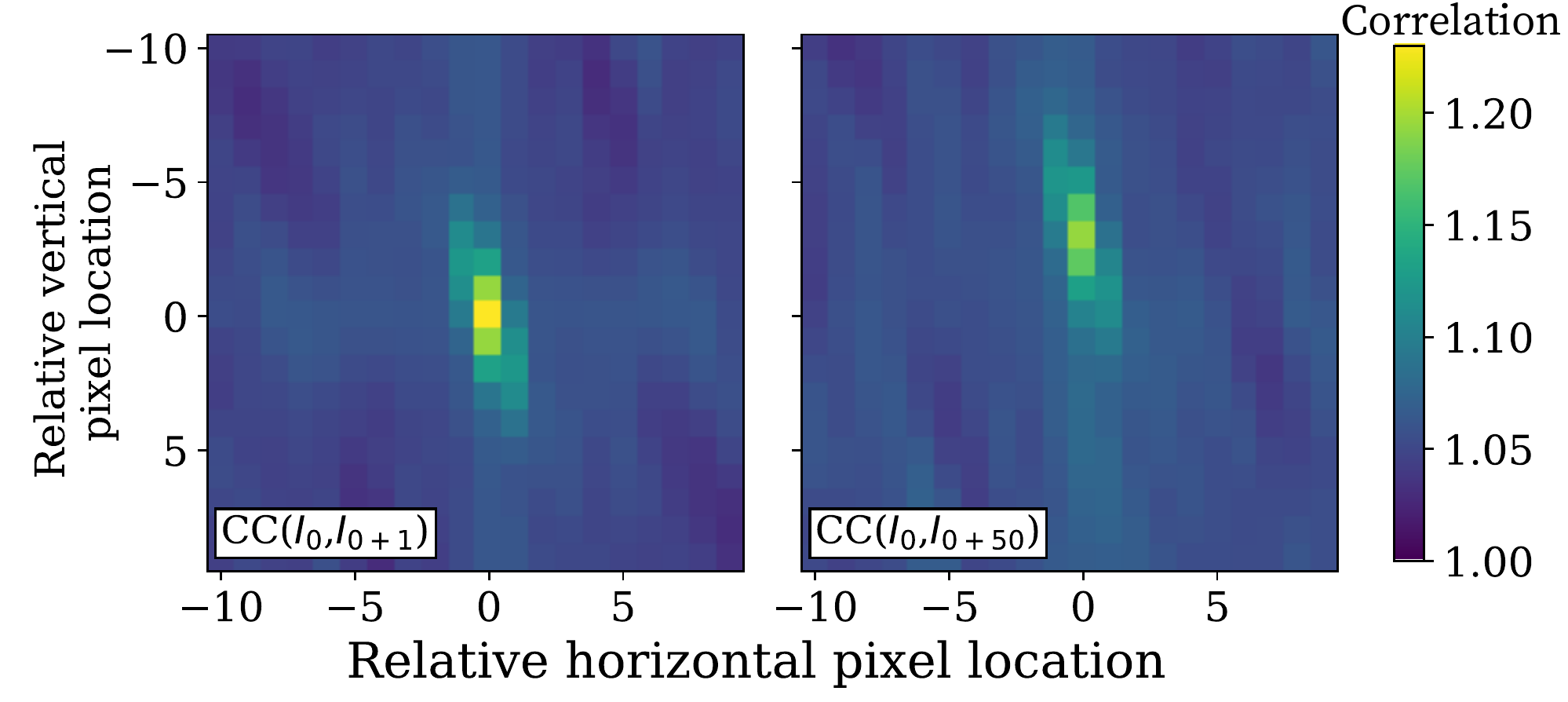}%

\vspace{4mm}

\includegraphics[width=3.1in]{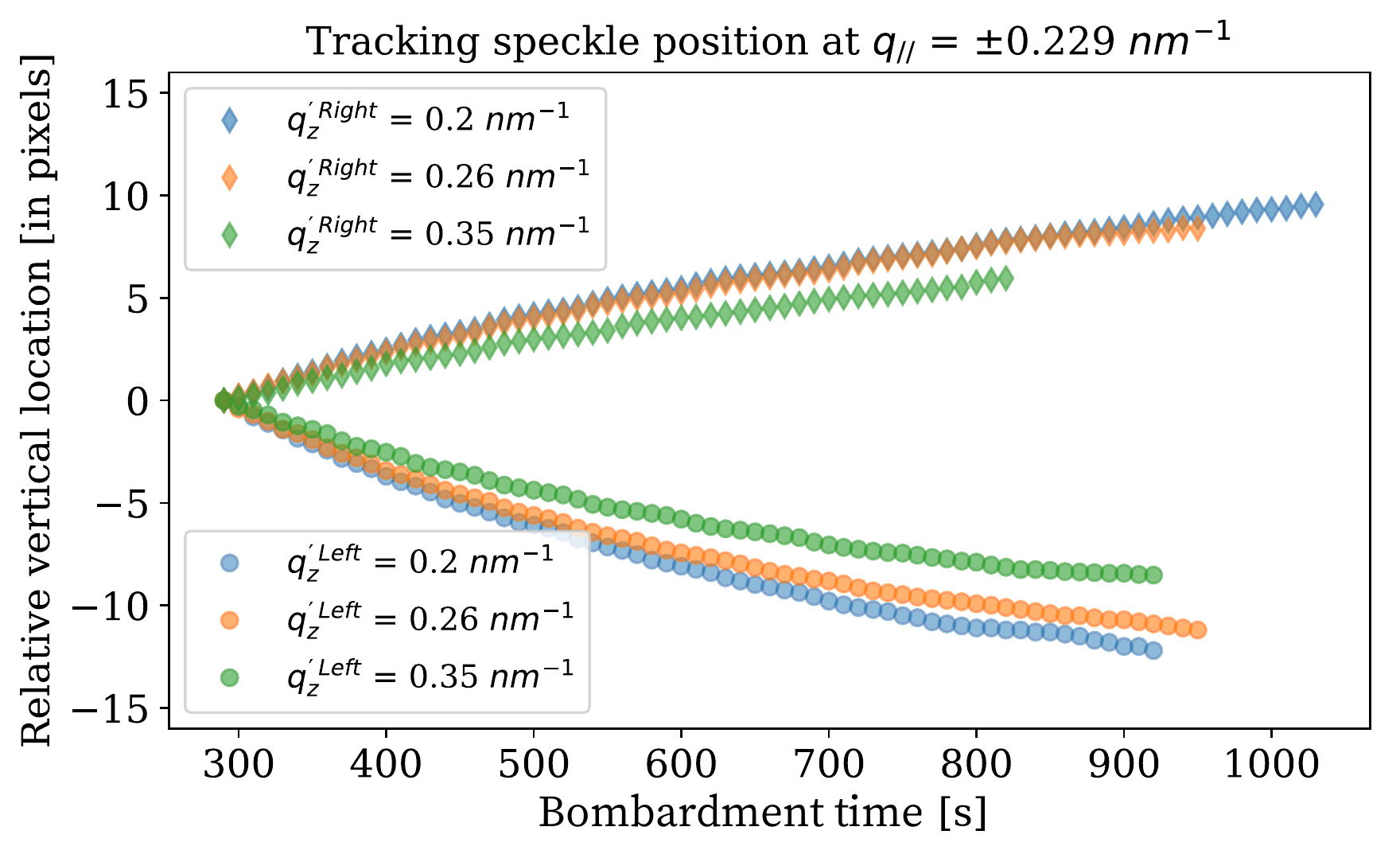}
\caption{\label{fig:Kr_cross_corr} Top: Cross-correlation analysis to track speckle movement. CC($I_{0}$,$I_{0+m}$) is the cross-correlation between the intensity pattern at bombardment time $t = 290$ s in a region of interest, denoted $I_0$, and the intensity pattern of that region a time $m$ seconds later, denoted $I_{0+m}$. The change in position of the cross-correlation peak from its original relative position at the origin shows how much speckle has moved between the two times.  Bottom: Resulting time-dependence of speckle positions relative to an initial position at time 290 s.  The regions of interest analysed were at $\pm 0.229 \; \mathrm{nm}^{-1}$ and various $q_z'$ as noted. Note that the zero level corresponds to the starting position for each individual $q_z'$, and thus the absolute pixel value (not shown) varies from one $q_z'$ to another.}
\end{figure}

In order to track speckle motion with time, cross correlations within a region of interest on the detector at different times $t_1$ and $t_2$ were calculated:
\begin{equation}
\label{equ:cross_corr}
CC(\mathbf{q},I(t_1),I(t_2)) = \frac{\mathcal{F}^{-1}(\mathcal{F}(I(t_1))\odot\mathcal{F}(I(t_2))^\dagger)}{\mathcal{F}^{-1}(\mathcal{F}(I(t_1)))  \odot \mathcal{F}^{-1}(\mathcal{F}(I(t_2))^\dagger)}
\end{equation}
where $\odot$ indicates a point-wise operation, and $\mathcal{F}$ and $\mathcal{F}^{-1}$ indicate forward and inverse Fourier transforms. Two cross-correlation results are shown in the top of Fig. \ref{fig:Kr_cross_corr}, where the peak of cross-correlation moved spatially, indicating speckle movement. In order to track the motion with sub-pixel resolution, a three-point quadratic function was first used to fit the peaks: $f(x,y) = a(x-x_0)^2+b(y-y_0)^2+c$. For example, by choosing three points that include the peak in the horizontal axis to be $f_{-1} = f(x-1,y), f_0 = f(x,y), f_1 = f(x+1,y)$, we have $a=(f_{-1}+f_1-2f_0), x_0 = (f_{-1}-f_1)/(4a), c= f_0-a x_0^2$. Fitting the spatial correlation in both $x$- and $y$-directions, it is possible to track the correlation peak's $x_0$ and $y_0$ positions.

\begin{figure}
\includegraphics[width=3.1 in]{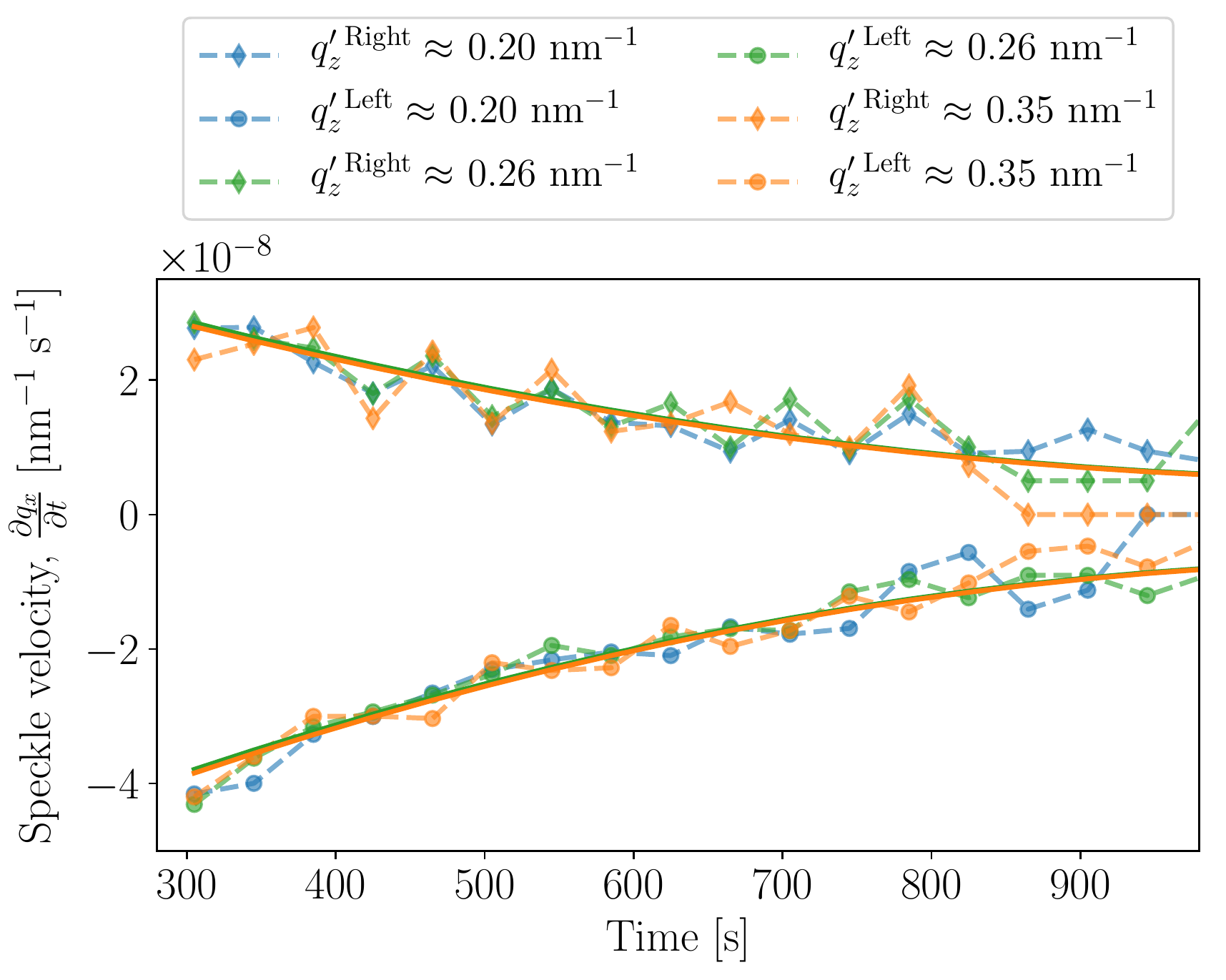}%
\caption{\label{fig:Kr_speckle_velocity} Speckle velocities $\frac{dq_x}{dt}$ as a function of time on the right side of the detector at $q_{||} = +0.229 \; \mathrm{nm}^{-1}$ and on the left side of the detector at $q_{||} = -0.229 \; \mathrm{nm}^{-1}$.  For each $q_{||}$, behavior is shown at three different $q_z^{\prime}$ values. The solid lines are from a simultaneous fit using the quadratic function of time discussed in the text.}
\end{figure}

\begin{figure}
\includegraphics[width=3.1 in]{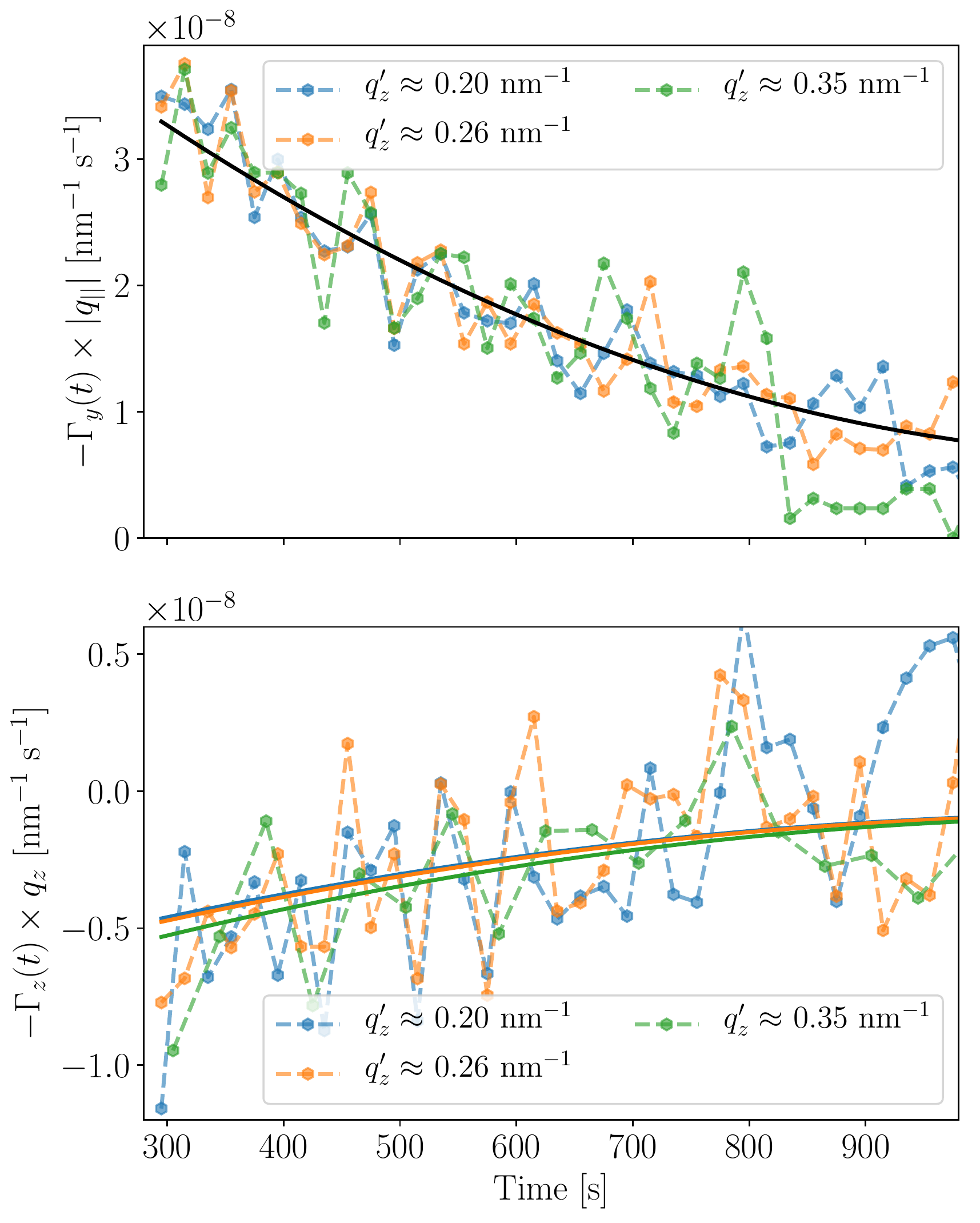}%
\caption{\label{fig:Kr_speckle_gammas}  The difference (top) and summation (bottom) of $dq_x/dt$ between the two sides of the detector (i.e. between $q_{||} = +0.229 \; \mathrm{nm}^{-1}$ and $q_{||} = -0.229 \; \mathrm{nm}^{-1}$) using Eq. \ref{equ:dqgammay}. The solid lines are from the single simultaneous fit using the quadratic function of time discussed in the text.}
\end{figure}

\begin{figure}
\includegraphics[width=3.1 in]{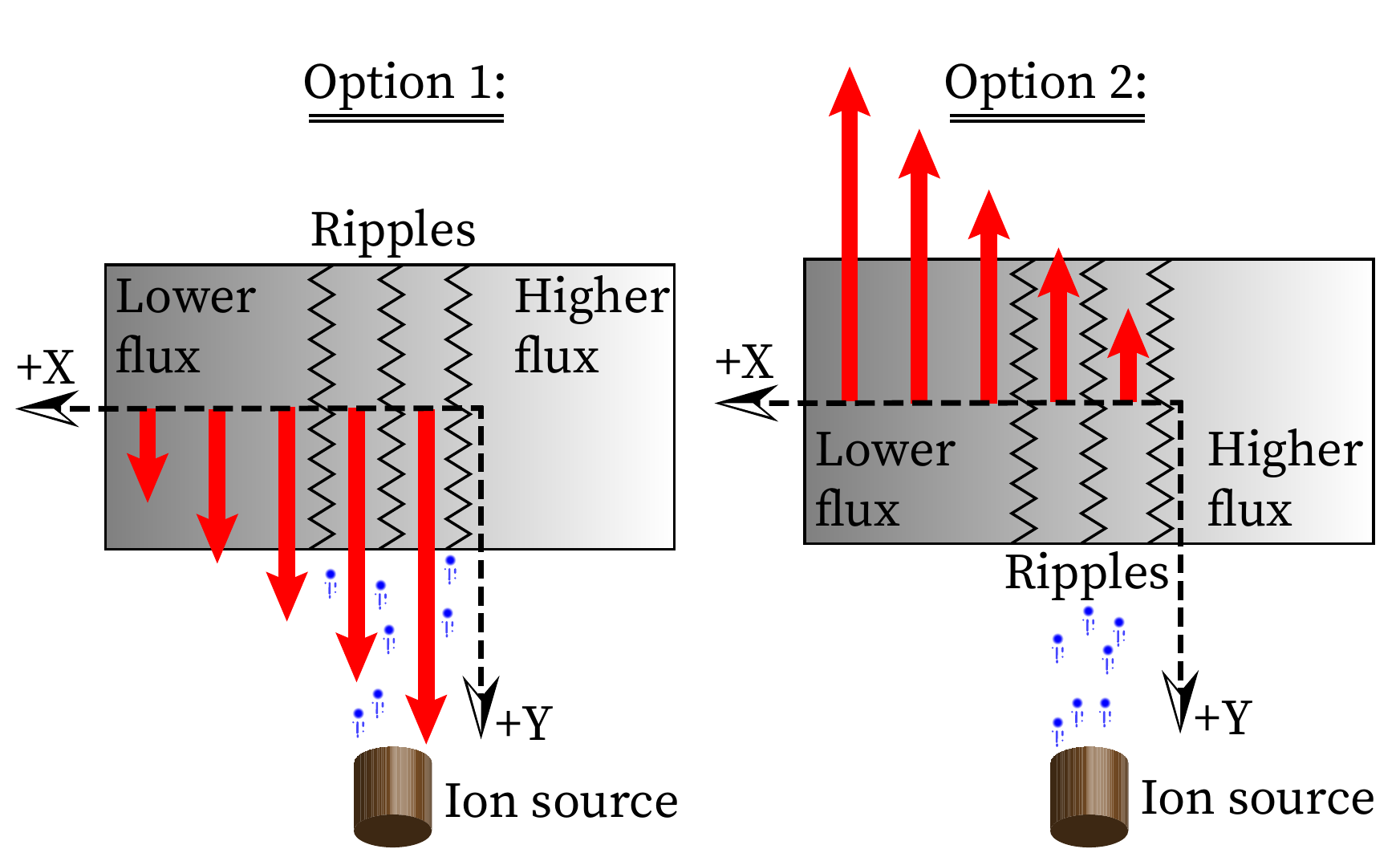}%
\caption{\label{fig:velocity_direction_schematic}Explanation of the determination of ripple velocity direction. Red vectors are ripple velocities with negative $\Gamma_y = \Delta v_y/\Delta x$. The measured positive $\Gamma_z$ value means that ion flux $F$ decreases with increasing $x$-coordinate as shown with the gradient in grey scale. Since flux $F \propto |v_y|$, only Option 1 is possible.}
\end{figure}

As shown in the bottom of Fig. \ref{fig:Kr_cross_corr}, speckles move vertically upward on the right side of the detector (i.e. for positive $q_{||}$) and vertically downward on the left side (i.e. for negative $q_{||}$). There is no horizontal movement. To determine the $\Gamma_y$ and $\Gamma_z$ parameters, the observed speckle motions were used to calculate $dq_x/dt$.  Since the $x$-component of the photon wavevector, $k_x$, is conserved across the material surface, calculation of $q_x$ using Eq. \ref{equ:wavenumber_conversion} simply uses the geometric scattering angles without refraction effects. Figure \ref{fig:Kr_speckle_velocity} shows that the magnitude of $dq_x/dt$ decreases steadily during the experiment and is nearly zero by the end.  The analysis presented here uses data at $q_{||} = \pm 0.229 \; \mathrm{nm}^{-1}$, but analyses were also performed at several other $q_{||}$ values and produced the same results.

To gain further insight, we take advantage of the structure of Eq. \ref{equ:dqx} to isolate the effects of the two different parameters $\Gamma_y(t)$ and $\Gamma_z(t)$ by adding and subtracting motion $dq_x/dt$ at pairs of detector points having equal $q_z$ but opposite $q_{||}$:
\begin{equation}
\begin{split}
\frac{1}{2} \left[ \frac{dq_x}{dt} (+|q_{||}|,q_z) - \frac{dq_x}{dt} (-|q_{||}|,q_z) \right] = -\Gamma_y(t) |q_{||}| \\
\frac{1}{2} \left[ \frac{dq_x}{dt} (+|q_{||}|,q_z) + \frac{dq_x}{dt} (-|q_{||}|,q_z) \right] = - \Gamma_z(t)\, q_{z}.
\label{equ:dqgammay}
\end{split}
\end{equation}
Figure \ref{fig:Kr_speckle_gammas} shows that the magnitude of the difference in $dq_x/dt$ between the two sides decreases steadily during the experiment, clearly indicating that the magnitude of the gradient velocity $\Gamma_y (t) = \Delta v_y(x,t)/\Delta x$ is decreasing with time.  The summed $dq_x/dt$ is noisier, but its magnitude is also decreasing throughout the experiment, suggesting that $\Gamma_z (t) = \Delta v_z(x,t)/\Delta x$ is also decreasing.  As seen in Eq. \ref{equ:velocity}, the ripple velocity is proportional to the ion flux so the most prosaic explanation for these observations is that the ion flux profile $F(x)$ is changing slightly over time, becoming more uniform.  This explanation suggests that $\Gamma_y \propto \Gamma_z$.

In fitting the observed temporal evolution of $dq_x/dt$ to Eq. \ref{equ:dqx}, the shape of the curves in Figs. \ref{fig:Kr_speckle_velocity} and \ref{fig:Kr_speckle_gammas} leads us to fit $\Gamma_y (t)$ as a quadratic function of time, i.e. $\Gamma_y(t) = a_0 + a_1 (t-290 \; \mathrm{s}) + a_2 (t-290 \; \mathrm{s})^2$.  This, in addition to the requirement that the two $\Gamma$'s are proportional, i.e. $\Gamma_z(t) = k \, \Gamma_y(t)$, gives a total of four fit parameters ($a_0, a_1, a_2$ and $k$) to simultaneously fit all of the speckle motion data.  The lines in Figs. \ref{fig:Kr_speckle_velocity} and \ref{fig:Kr_speckle_gammas} are the resulting fits.  The fit parameters are: $a_0 = -1.49 \times 10^{-7}\; \mathrm{s}^{-1}$, $a_1 = 2.87 \times 10^{-10} \; \mathrm{s}^{-2}$, $a_2 = -1.67 \times 10^{-13}\; \mathrm{s}^{-3}$ and $k=-7.63 \times 10^{-2}$.  

At the earliest time shown, 290 s after the beginning of bombardment, $\Gamma_z = 1.14 \times 10^{-8} \; \mathrm{s}^{-1}$.  If the gradient in erosion rate is indeed due to a slight inhomogeneity in ion flux, then the flux inhomogeneity can be calculated from $\Gamma_z = \Delta v_z(x)/\Delta x = -(\Delta F(x) /\Delta x) \Omega Y(\theta) \cos\theta$.  The sputter yield at 65$^{\circ}$ as calculated by SPTrimSD is $Y(\theta=65^{\circ}) = 4.36$, which then gives an ion flux gradient $\Delta F(x)/\Delta x = -3.1 \times 10^{13}$ (ions/cm$^2$s)/mm.  Given the average incident flux $1 \times 10^{15} \; \mathrm{ions/cm}^2\mathrm{s}$, this corresponds to $\sim$ 3\% ion flux gradient in 1 mm.

The negative sign of $\Gamma_y$ shows that the negative gradient in ion flux leads to a negative change in $v_y$ with increasing $x$-coordinate.  As shown in Fig. \ref{fig:velocity_direction_schematic}, this is only possible if the ion flux causes the ripples to move in the positive $y$-direction, i.e. in the direction heading into the oncoming ion beam.  Thus the the ion bombardment angle $\theta = 65^{\circ}$ is found to be \textit{below} the transition angle $\theta_c$.

Mathematically, the ratio of ripple velocity to flux is found to be:
\begin{equation}
\label{equ:deltas}
\begin{split}
\frac{v_y(x)}{F(x)} = \frac{\Delta v_y(x)/\Delta x}{\Delta F(x)/\Delta x} = \left( \frac{\Gamma_y}{\Gamma_z} \right) \Omega Y(\theta) \cos(\theta)\\
= \frac{\Omega Y(\theta) \cos(\theta)}{k} = 4.8 \times 10^{-15} \left( \frac{\mathrm{nm}}{\mathrm{s}} \right) / 
\left( \frac{\mathrm{ion}}{\mathrm{cm}^2 \mathrm{s}} \right)
\end{split}
\end{equation}
where the fit value of $k$ from above and $Y(\theta=65^{\circ}) = 4.36$ from SDTrimSP are used.  With the average flux of  $1 \times 10^{15} \; \mathrm{ions/cm}^2\mathrm{s}$, Eq. \ref{equ:deltas} yields a ripple velocity of 4.8 nm/s.

\section{Discussion and Conclusions}

Comparison of the present results from Kr$^+$ nanopatterning of Si with those from Ar$^+$ nanopatterning presented in Part I is instructive.  As noted in Sect. \ref{sec:early_kinetics}, the ion enhanced viscous flow relaxation in the two cases is essentially equal.  Thus the change in ion mass at constant energy does not seem to effect the activation of such relaxation mechanisms.  The curvature-dependent term, however, increases in magnitude in going from Ar$^+$ to Kr$^+$ in a manner consistent with expectations for increased lateral mass redistributive processes for the heavier ion with its larger momentum.

As found also for Ar$^+$ nanopatterning in Part I of this work, there are strong relationships between the correlation time $\tau(q)$ and the surface structure as measured by the scattering pattern.  Figure \ref{fig:Kr_g2} shows that, near the peak wavenumbers $\pm \; q_0$, the scattering intensity initially grows much more rapidly than does the correlation time $\tau$, but that at later times the intensity grows only slowly while $\tau$ continues to grow significantly.

Beyond the initial ripple growth process, the relaxation of fluctuations is found to be compressed exponential near the peak wavenumber $q_0$ and stretched exponential at higher wavenumbers $|q_{||}|$.  As discussed Ref. \cite{myint2021gennes} and in Part I of the study, compressed exponential behavior at short times $\Delta t$ in both soft materials \cite{cipelletti2005slow} and metallic glasses \cite{ruta2012atomic} has been attributed to collective ballistic flow of local structures due to internal stress relaxation. Some theoretical approaches to understanding ion beam nanopatterning use fluid dynamic models with stress relaxation as a driving force \cite{castro2012hydrodynamic,castro2012stress,norris2012stress,moreno2015nonuniversality,munoz2019stress}. These might provide a direct connection between the compressed exponential behavior of ion beam nanopatterning observed here and that observed in glasses. 

A significant step forward in the present work has been the demonstration of using speckle motion to reveal detailed information about the sputter erosion rate and ripple velocity.  The approach developed here is a real-time measurement applicable to surfaces with even short ripple wavelengths, and offers the possibility of wide applicability. Higher accuracy could likely be achieved by using a larger flux gradient.  

For the particular case of self-organized Si rippling by 1 keV Kr$^+$ bombardment, this study finds that the ion incidence angle of 65$^{\circ}$ is below the transition angle of ripple velocity $\theta_c$. Calculation of Eq. \ref{equ:velocity} using $Y(\theta)$ values from SDTrimSP gives a predicted transition angle $\theta_c \approx 58^{\circ}$, in conflict with the measurement.  In addition, for angles below $\theta_c$, use of SDTrimSP $Y(\theta)$ values in Eq. \ref{equ:velocity} gives a maximum ripple velocity that is approximately an order of magnitude smaller than the velocity measured here.  In light of these results it is noteworthy that the ripple motion measurements of Hofs{\"a}ss \textit{et al.} \cite{hofsass2013propagation} for 10 keV Xe$^+$ patterning of Si, found that the experimental $\theta_c$ was also slightly higher than predicted by Eq. \ref{equ:velocity} and found a speed at $\theta$ = 62$^{\circ}$ that was larger than the maximum predicted for $\theta < \theta_c$.  

In some theories, stress effects play a significant role in determining the ripple velocity \cite{moreno2015stress}.  It may be possible that their inclusion would produce better agreement with experiment, though the relevant example calculations of Ref. \cite{moreno2015stress} would seem to decrease, rather than increase, the transition angle $\theta_c$.  It would be quite interesting to investigate this further with measurements at multiple ion incidence angles.  In addition, it's believed that the ripple velocity may change, and possibly reverse sign, as sawtooth patterns evolve at later bombardment times than those examined in this study \cite{pearson2014theory}.  The real-time nature of the speckle motion technique developed here would be ideal to investigate the existence and nature of such behavior.

\begin{acknowledgments}
We thank Mark Sutton for many useful discussions and for use of his speckle tracking program. We also thank Andreas Mutzke for providing the SDTrimSP simulation program and S. Norris for help with the PyCraters library.  The component of this work at BU was partly supported by the National Science Foundation (NSF) under Grant No. DMR-1709380. At UVM, X.Z. and R.H. were partly supported by the U.S. Department of Energy (DOE) Office of Science under Grant No. DE-SC0017802. Experiments were done at the Coherent Hard X-ray (CHX) beamline at National Synchrotron Light Source II (NSLS-II), a U.S. Department of Energy (DOE) Office of Science User Facility operated for the DOE Office of Science by Brookhaven National Laboratory under Contract No. DE-SC0012704. The custom UHV sample holder, designed by P.M. and K.F.L, was built at Scientific Instrumentation Facility (SIF) at Boston university. For the AFM images, Bruker Dimension 3000 Atomic Force Microscope at Precision Measurement Laboratory at the Boston University Photonics Center was utilized.
\end{acknowledgments}

\bibliography{ionpatterning_references.bib}

\end{document}